\documentclass[prb,amsmath,twocolumn,showpacs,superscriptaddress]{revtex4}
\usepackage{color}
\usepackage{graphics}
\usepackage{subfigure}
\usepackage{ulem}
\usepackage{times}
\usepackage{graphicx}
\usepackage{amsmath}
\usepackage{amsbsy}
\usepackage{dcolumn}
\usepackage{bm}
\usepackage{epsfig}
\usepackage{color}
\usepackage{mathtools}
\usepackage{dsfont}

\begin{document}

\title{Manifestation of nematic degrees of freedom in the Raman response function of iron pnictides}

\author{U. Karahasanovic}
\affiliation{Institut f\"ur Theorie der Kondensierten Materie, Karlsruher Institut f\"ur Technologie, DE-76131 Karlsruhe, Germany}
\affiliation{Institut f\"ur Festk\"orperphysik, Karlsruher Institut f\"ur Technologie, DE-76131 Karlsruhe, Germany}
\author{F. Kretzschmar}
\affiliation{Walther Meissner Institut, Bayerische Akademie der Wissenschaften, 85748 Garching, Germany}
\affiliation{Fakult\"at f\"ur Physik E23, Technische Universit\"at M\"unchen, 85748 Garching, Germany}
\author{T. B\"ohm}
\affiliation{Walther Meissner Institut, Bayerische Akademie der Wissenschaften, 85748 Garching, Germany}
\affiliation{Fakult\"at f\"ur Physik E23, Technische Universit\"at M\"unchen, 85748 Garching, Germany}
\author{R. Hackl}
\affiliation{Walther Meissner Institut, Bayerische Akademie der Wissenschaften, 85748 Garching, Germany}
\author{I. Paul}
\affiliation{Laboratoire Mat\'eriaux et Ph\'enom\`enes Quantiques, UMR 7162 CNRS, Universit\'e
Paris Diderot, Bat. Condorcet 75205 Paris Cedex 13, France}
\author{Y. Gallais}
\affiliation{Laboratoire Mat\'eriaux et Ph\'enom\`enes Quantiques, UMR 7162 CNRS, Universit\'e
Paris Diderot, Bat. Condorcet 75205 Paris Cedex 13, France}
\author{J. Schmalian}
\affiliation{Institut f\"ur Theorie der Kondensierten Materie, Karlsruher Institut f\"ur Technologie, DE-76131 Karlsruhe, Germany}
\affiliation{Institut f\"ur Festk\"orperphysik, Karlsruher Institut f\"ur Technologie, DE-76131 Karlsruhe, Germany}

\date{\today}

\begin{abstract}

We establish a relation between the Raman response function in the $B_{1g}$ channel and the electronic contribution to the nematic susceptibility within the spin-driven approach to electron nematicity  of the iron based superconductors. The spin-driven nematic phase, characterized by the broken $C_4$ symmetry, but unbroken $O(3$) spin-rotational symmetry, is generated by the presence of magnetic fluctuations associated with the striped phase. It occurs as a separate phase between $T_m$ and $T_s$ in systems where the structural  and magnetic phase transitions are separated.  Detecting the presence of nematic degrees of freedom in iron-based superconductors is a difficult task, since it involves measuring higher order spin correlation functions. We show that the nematic degrees of freedom manifest themselves in the experimentally measurable Raman response function. We calculate the Raman response function in the tetragonal phase in the large $N$ limit by considering higher-order Aslamazov-Larkin type of diagrams. They are characterized by a series of inserted quartic paramagnon couplings mediated by electronic excitations that resemble the nematic coupling constant of the theory. These diagrams effectively account for collisions between spin fluctuations. By summing an infinite number of such higher order diagrams, we demonstrate that the electronic Raman response function shows a clear maximum at the structural phase transition in the $B_{1g}$ channel. Hence, the Raman response function can be used to probe nematic degrees of freedom.

\end{abstract}

\pacs{74.25.nd, 74.70.Xa, 74.20.Mn, 74.25.Ha}

\maketitle
\section{Introduction}

Iron-based superconductors show rich phase diagrams, with the high-temperature superconducting dome being in the close proximity to an antiferromagnetic striped phase \cite{review} that sets in at a temperature $T_m$. In addition, a structural phase transition at $T_s$, from the  high-temperature tetragonal phase into an orthorhombic phase, has been shown to closely follow the magnetic transition \cite{Birgeneau11, Kim11, matsuda_t, nematic_review}, i.e.: $T_s\geq T_m$ . It was proposed that spin-fluctuations, associated with the striped phase, lead to emergent electronic nematic degrees of freedom at higher temperatures. \cite{Fernandes12, naturereview, nematic_review, Xu08, Fang08} These electronic nematic degrees of freedom then couple to the lattice and induce the structural phase transition to the ortorhombic phase. \cite{Qi09,Fernandes2010,Cano10}

There is a mounting evidence for an electronic nematic state: resistivity-anisotropy measurements \cite{Tanatar10, Chu2012} and the measurement of the elastoresistance, \cite{Chu10} the observed anisotropies in thermopower, \cite{Jiang2013} optical conductivity, \cite{Dusza2011,Nakajima2011} torque magnetometry, \cite{matsuda_t} and in STM measurements. \cite{Rosenthal13} Measurements of the elastic constants showed that the shear modulus strongly softens in the high temperature tetragonal phase.\cite{Fernandes13_shear, Fernandes2010, Kontani1, Kontani2, Anna1} A theoretical analysis \cite{Fernandes2010} based upon nematic fluctuations due to a strong magneto-elastic coupling showed that the inverse shear modulus is proportional to the susceptibility of the nematic order parameter $\chi_{\mathrm{nem}}$, which diverges at the structural phase transition, explaining its softening. The most direct evidence for the magnetic origin of nematicity so far is the scaling of the shear modulus and the NMR spin-lattice relaxation rate, seen in iron-pnictides. \cite{Fernandes13_shear} An interesting open issue in this context is the lack of such scaling behavior in iron-chalcagonides. \cite{Anna2}

A relation between nematicity and the Raman response of iron based superconductors was already studied in Ref. \onlinecite{Gallais} where the Kramers-Kronig transform of the Raman response was compared with the shear modulus, as well as in Ref. \onlinecite{Blumberg1, Blumberg2}. 
Here, we demonstrate, based on an explicit microscopic theory that i) there is no enhancement of the electronic Raman response function in the $B_{2g}$ channel upon lowering the temperature, ii) that the Raman response function develops a pronounced peak at the structural phase transition in the $B_{1g}$ symmetry, and iii) that there is some response in the $A_{1g}$ channel, which weakens as the temperature is lowered towards the structural transition temperature. 

We start from the spin-driven scenario for the nematic phase, in which magnetic fluctuations stabilize a nematic phase, characterized by the broken C$_4$ symmetry. The Raman response function measures the electronic density-density correlator weighted by appropriate form factors. Since electrons interact with spin fluctuations, the latter will manifest themselves in the Raman response function in the form of  corrections to the electron self energy and the Raman vertex, formally expressed in terms of Aslamazov-Larkin diagrams. \cite{Caprara05} We show that the leading order Aslamazov-Larkin (AL) diagram supports only the $A_{1g}$ and the $B_{1g}$ symmetry, but not the $B_{2g}$ symmetry, which explains the lack of enhancement of the Raman response signal as one approaches the structural transition in the $B_{2g}$ channel, as seen in experiments. \cite{Rudi, Gallais} However, this leading order approach cannot account for the rapid increase in the amplitude of the Raman response function in the $B_{1g}$ channel, as one approaches the structural transition, as seen in the experiments of Refs. \onlinecite{Rudi, Gallais}. Instead it would predict a similar increase only at the magnetic phase transition.  Therefore, we go beyond the leading order approximation, and take into account collisions between spin fluctuations that become more and more important as one approaches the  nematic / structural transition. Our approach is based on the exact same collisions between spin-fluctuations that led to the emergence of spin-induced nematicity in the first place. Formally this is accomplished by
inserting a series of quartic paramagnon couplings, mediated by electronic excitations, into the Raman response function. Such quartic couplings contain a product of four fermionic Green's functions and give rise to a peak of the electronic Raman response function at the structural phase transition in the $B_{1g}$ channel. On the other hand, if we re-sum such higher order AL diagrams in the $A_{1g}$ channel, this will lead to the suppression of the Raman response in the aforementioned channel.

Here we demonstrate that the low frequency Raman response in the $B_{1g}$ channel is given by
\begin{equation}
R_{B_{1g}}\left( \omega \right)= \frac{R_0(\omega)}{1-\tilde g\int_q \chi_{q}^2},
\label{eq:Reqchi}
\end{equation}
where $\omega$ is the frequency difference between incoming and outgoing photons and $q$ the multi-index for momentum and frequency. $R_0(\omega)$ stands for the leading order Aslamazov-Larkin diagram, $\chi_{q}$ is the magnetic susceptibility, and $\tilde g$ the nematic coupling constant of the theory.
On the other hand, the susceptibility of the nematic order parameter of our model, in the large $N$ limit is given by
\begin{equation}
\chi_{\rm nem}=\frac{\int_q \chi^2_{q}}{1-g_{\rm stat} \int_q \chi^2_{q}},
\end{equation}
where in a purely electronic theory $\tilde g=g_{\rm stat}$. In a purely electronic theory, this would then lead to the divergence of the Raman response function at the structural phase transition. However, one needs to include the effect of the lattice dynamics \cite{Kontani1} in order to analyze this problem. We do so by introducing nemato-elastic coupling and find that, in this case, $ g_{\rm stat}=\tilde g + \frac{\gamma^2_{\rm el}}{c_{\rm s}^{0}}$ is shifted. \cite{Indranil} Here $\gamma_{\rm el}$ is the elasto-nematic coupling constant, and $c_{\rm s}^{0}$ the bare value of the orthorhombic elastic constant. 
We show that when magnetic and structural phase transitions are split \cite{Kim11, Birgeneau11, matsuda_t} this leads to a maximum of the amplitude of the electronic Raman response function in the $B_{1g}$ channel at the structural phase transition, in agreement with the recent experiments. \cite{Rudi} 
The Raman response function could then be used to probe the dynamic excitation spectrum of the nematic degrees of freedom, similar to inelastic neutron scattering that probes the dynamic spin excitation spectrum.

The paper is organized as follows. In Sec. \ref{microscopics} we present the microscopic model for the spin-driven nematic phase. We calculate the effective action and analyze it in the large-$N$ limit, where $N^2-1$ is the number of components of the collective paramagnon field. Following Ref. \onlinecite{Fernandes12}, we derive the condition for the susceptibility of the nematic order parameter to diverge. In Sec. \ref{Raman} we then show how to calculate the Raman response function using a diagrammatic approach. We first calculate the leading order Aslamazov Larkin diagram, and show that there is no response in the $B_{2g}$ channel, and a finite response in the $B_{1g}$ and the $A_{1g}$ channels. We then calculate higher order diagrams that take into account collisions between spin-fluctuations. Finally, after summing an infinite number of these higher-order diagrams within a controlled $1/N$ expansion, we show i) that the maximum of the Raman response function in the $B_{1g}$ channel occurs when the nematic susceptibility diverges, i.e. at the structural phase transition, and ii) that the amplitude of the Raman response function in the $A_{1g}$ response gets suppressed. We present our conclusions in Sec. \ref{conclusion}.

\section{Microscopic model: spin driven nematicity}
\label{microscopics}

Two different approaches have been proposed in order to explain the origin of nematic phase in pnictides and its relation to the magnetic phase -- the orbital scenario \cite{Phillips11, Applegate11, Dagotto13, w_ku10, kruger1, kruger2} and the spin-driven nematic scenario. \cite{Fernandes12, naturereview, nematic_review} For a discussion of these approaches see for example Ref. \onlinecite{naturereview}. Here we follow the approach of a spin-driven nematic state.  In this scenario, the nematic phase is stabilized by magnetic fluctuations that are associated with the stripe density wave (SDW) phase. The order parameter of the SDW state\cite{Dai_review} can be characterized by an $O(3) \times Z_2$ manifold \cite{Kivelson, Sachdev} -- $O(3)$ is the spin-rotational symmetry and $Z_2$ a discrete symmetry associated with the choice of the ordering wave-vector, ${\mathbf{Q}}_{X}=(\pi,0)$ or ${\mathbf{Q}}_{Y}=(0,\pi)$. Let the two order parameters associated with these two ordering wave vectors be ${\mathbf{\Delta}}_{X}$ and ${\mathbf{\Delta}}_{Y}$ respectively. The SDW state is characterized by broken $O(3)$ and $Z_2$ symmetries. On the mean-field level the breaking of $Z_2$ and $O(3)$ symmetry occurs simultaneously. However, when one includes fluctuations, these transitions can be split. In case of joint transitions, they are usually both first order transitions.\cite{Fernandes12} The criterion for breaking the discrete $Z_2$ symmetry via a second order transition is a threshold value of the magnetic correlation length $\xi$. Decreasing the temperature leads to an increase of $\xi$. Before the correlation length diverges at the magnetic phase transition temperature, the threshold value will be reached and spin-driven nematicity sets in. This naturally explains why the magnetic and structural phase boundaries are correlated and leads to an intermediate phase with $Z_2$ symmetry breaking without $O(3)$ symmetry breaking. This intermediate state  is the nematic phase in the pnictides. It is characterized by unequal strength of the magnetic fluctuations associated with the ordering wave vectors ${\mathbf{Q}}_{X}$ and ${\mathbf{Q}}_{Y}$ : $\langle {\mathbf{\Delta}}_{X}^2-{\mathbf{\Delta}}_{Y}^2 \rangle \neq 0$, but no long range magnetic order, $\langle {\mathbf{\Delta}}_{X,Y} \rangle =0$.

\begin{figure}
\begin{centering}
\includegraphics[width=1\columnwidth]{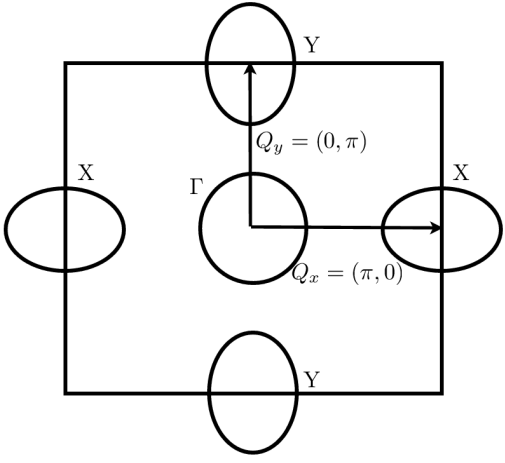} 
\par\end{centering}
\caption{Band structure: the model consists of the central hole-like
$\Gamma$ band, and the electron-like $X$ and $Y$ bands, shifted
by ${\mathbf{Q}}_{X}=(\pi,0)$ and ${\mathbf{Q}}_{Y}=(0,\pi)$, respectively.}
\label{bandstructure}
\end{figure}
 
In what follows we will summarize the steps of Ref. \onlinecite{Fernandes12} and outline the mathematical model the for spin-driven nematic phase. We start from a simplified itinerant model where we include the bands near the $\Gamma-$point and the $X-$ and $Y-$ points in the Brillouin zone. For our main result no explicit knowledge of the detailed parametrization of the band structure is necessary, except for the fact that the band-structure is not perfectly nested. However, in order to obtain explicit numerical results we use the simplified model of Ref. \onlinecite{Fernandes12}.  We consider parabolic dispersions with

\begin{eqnarray}
\epsilon_{\Gamma,\mathbf{k}} & = & \epsilon_{0}-\frac{k^{2}}{2m}-\mu ,
\nonumber \\
\epsilon_{X,\mathbf{k+Q_{X}}} & = & -\epsilon_{0}+\frac{k_{x}^{2}}{2m_{x}}+\frac{k_{y}^{2}}{2m_{y}}-\mu ,
\nonumber \\
\epsilon_{Y,\mathbf{k+Q_{Y}}} & = & -\epsilon_{0}+\frac{k_{x}^{2}}{2m_{y}}+\frac{k_{y}^{2}}{2m_{x}}-\mu ,
\end{eqnarray}
where $m_{i}$ are the band masses, $\epsilon_{0}$ is the
offset energy, and $\mu$ denotes the chemical potential. The corresponding Fermi surfaces are shown in Fig.\ref{bandstructure}.

In order to study the established stripe magnetic phase, we consider the Hamiltonian that contains the interactions in the spin channel with momenta near $\mathbf{Q}_{X}$ and $\mathbf{Q}_{Y}$:
\begin{eqnarray}
\mathcal{H} & = & \mathcal{H}_{0}+\mathcal{H}_{\mathrm{int}},
\nonumber \\
\mathcal{H}_{0} & = & \sum\limits _{i,\mathbf{k}}\epsilon_{i,\mathbf{k}}c_{i,\mathbf{k}\alpha}^{\dagger}c_{i,\mathbf{k}\alpha},
\nonumber \\
\mathcal{H}_{\mathrm{int}} & = &-\frac{1}{2}u_{\mathrm{s}}\sum\limits _{i,\mathbf{q}}\mathbf{s}_{i,\mathbf{q}}\cdot\mathbf{s}_{i,-\mathbf{q}}\, .
\label{Hamiltonian}
\end{eqnarray}
Here, $c_{i,\mathbf{k}\alpha}^{\dagger}$ is the creation operator of an electron with momentum $\mathbf{k}$, spin $\alpha$ and in the band $i$. The spin operator is given by 
\begin{equation}
\mathbf{s}_{i,\mathbf{q}}=\sum \limits_{k}c_{\Gamma,\mathbf{k+q}\alpha}^{\dagger}\boldsymbol{\lambda}_{\alpha\beta}c_{i,\mathbf{k}\beta},
\end{equation}
 where $\boldsymbol{\lambda}_{\alpha\beta}$ denotes the $N^2-1$ component vector of the generators of the $SU(N)$ algebra. In the case $N=2$ it holds $\boldsymbol{\lambda}_{\alpha\beta}=\frac{1}{2}\boldsymbol{\sigma}_{\alpha\beta}$ with vector of the Pauli matrixes $\boldsymbol{\sigma}$. $u_{\mathrm{s}}$ is the coupling in the spin channel, which can be expressed in terms of density-density and pair-hopping interactions between hole and electron pockets. \cite{Chubukov08}

The partition function is given by
\begin{equation}
Z=\int dc_{i,\mathbf{k}}dc_{i,\mathbf{k}}^{\dagger}\mathrm{e}^{-\beta\mathcal{H}},
\label{Z}
\end{equation}
where $\beta=T^{-1}$ is the inverse temperature.
Since, the interaction Hamiltonian is quadratic in the fermionic spin $\mathbf{s}_{i,\mathbf{q}}$, we can decouple it using a Hubbard-Stratonovich decoupling in the spin-channel. This way, we eliminate the quartic interaction between fermions at the expense of a functional integral over two additional bosonic fields $\boldsymbol{\Delta}_{X}$ and
$\boldsymbol{\Delta}_{Y}$, with $N^2-1$ components. The bosonic fields couple linearly to the corresponding fermionic spin densities. After introducing the spinor
\begin{equation}
\Psi_{\mathbf{k}}^{\dagger}=\left(\begin{array}{ccc}
c_{\Gamma,\mathbf{k}\alpha}^{\dagger} & c_{X,\mathbf{k}\alpha}^{\dagger} & c_{Y,\mathbf{k}\alpha}^{\dagger}  \end{array}\right),
\label{nambu}
\end{equation}
where $\alpha$ denotes every possible value of the $SU(N)$ spin index, we can write the partition
function as:
\begin{equation}
Z=\int d\Delta_{i}d\Psi\mathrm{e}^{-S\left[\Psi,\Delta_{i}\right]},
\label{Z}
\end{equation}
with the action:
\begin{equation}
S\left[\Psi,\Delta_{i}\right]=-\int_{k}\Psi_{k}^{\dagger}\mathcal{G}_{\Delta, k}^{-1}\Psi_{k}^{}+\frac{2}{u_{\mathrm{s}}}\int_{x}\left(\boldsymbol{\Delta}_{X}^{2}+\boldsymbol{\Delta}_{Y}^{2}\right).
\label{S}
\end{equation}
Here, the matrix of the inverse Green's
function $\mathcal{G}_{k}^{-1}$ is given by:
\begin{equation}
\mathcal{G}_{\Delta,k}^{-1}=\mathcal{G}_{0,k}^{-1}-\mathcal{V}_{\Delta},
\label{G}
\end{equation}
with the bare term:
\begin{equation}
\mathcal{G}_{0,k}=\left(\begin{array}{ccc}
\hat{G}_{\Gamma,k} & 0 & 0\\
0 & \hat{G}_{X,k} & 0\\
0 & 0 & \hat{G}_{Y,k}\end{array}\right),
\label{G0}
\end{equation}
 and the interacting term:
\begin{equation}
\mathcal{V}_{\Delta}=\left(\begin{array}{ccc}
0 & -\boldsymbol{\Delta}_{X} \cdot \boldsymbol{\lambda} & -\boldsymbol{\Delta}_{Y} \cdot \boldsymbol{\lambda}\\
-\boldsymbol{\Delta}_{X} \cdot \boldsymbol{\lambda} & 0 & 0\\
-\boldsymbol{\Delta}_{Y} \cdot \boldsymbol{\lambda}& 0 & 0\end{array}\right).
\label{V_delta}
\end{equation}
 $\hat{G}_{i,k}=G_{i,k}\hat{1} $ with  $G_{i,k}^{-1}=i\omega_{n}-\epsilon_{i,\mathbf{k}}$ and $N \times N$ unit matrix $\hat{1} $.
We invert the matrix equation (\ref{G}) by expanding the geometric series and  obtain the following expression for $\mathcal{G}_{\Delta}$ that we will use later-on:  
\begin{eqnarray}
\mathcal{G}_{\Delta}=\sum_{n=0}^{\infty} \left ( \mathcal{G}_0 \mathcal{V}_{\Delta} \right )^n \mathcal{G}_{0}.
\label{invGdelta}
\end{eqnarray}

\subsection{Effective action in the large-$N$ expansion}
\label{eff action}
In this section, we first show how to obtain the Ginzburg-Landau expansion of the effective action in powers of the spin fluctuation fields $\boldsymbol{\Delta}_{X,Y}$ in the limit of large $N$, in the spirit similar to that of Ref. \onlinecite{Fernandes12} where only $N=2$ was considered. Next, we re-formulate this effective action in terms of the collective nematic Ising variable $\phi$, and analyze the equation of state for $\phi$. We deduce the condition for the onset of the nematic phase by examining the susceptibility of the nematic order parameter.
We begin by integrating out the fermionic degrees of freedom from Eq. (\ref{Z}). It follows:  
\begin{equation}
Z  =  \int d\Delta_{i}\mathrm{e}^{-S_{\mathrm{eff}}\left[\boldsymbol{\Delta}_{X},\boldsymbol{\Delta}_{Y} \right]}
\end{equation}
with action:
\begin{eqnarray}
S_{\mathrm{eff}}\left[\boldsymbol{\Delta}_{X},\boldsymbol{\Delta}_{Y}\right] &=&-\mathrm{Tr}\ln\left(1-\mathcal{G}_{0}\mathcal{V}_{\Delta}\right)
\nonumber \\
& &+\frac{2}{u_{\mathrm{s}}}\int_{x}\left(\boldsymbol{\Delta}_{X}+\boldsymbol{\Delta}_{Y}^{2}\right)\label{effective_S1}.
\end{eqnarray}
Here, $\mathrm{Tr}\left(\cdots\right)$ refers to sum over momentum,
frequency, spin, and band indices. We further expand in powers of $\boldsymbol{\Delta}_{X,Y}$ to obtain:
\begin{eqnarray}
S_{\mathrm{eff}}\left[\boldsymbol{\Delta}_{X},\boldsymbol{\Delta}_{Y}\right]  &=& \frac{1}{2}\mathrm{Tr}\left(\mathcal{G}_{0,k}\mathcal{V}_{\Delta}\right)^{2}+\frac{1}{4}\mathrm{Tr}\left(\mathcal{G}_{0,k}\mathcal{V}_{\Delta}\right)^{4}\nonumber \\
 & &+ \frac{2}{u_{\mathrm{s}}}\int_{x}\left(\boldsymbol{\Delta}_{X}^{2}+\boldsymbol{\Delta}_{Y}^{2}\right).
 \label{effective_S2}
 \end{eqnarray}
After using a series of identities for the generators of the $SU(N)$ algebra, needed to evaluate the above traces  (for details see \ref{eff_action_derivation}), we arrive at the following effective action in the large $N$ limit:
\begin{equation}
S_{\mathrm{eff}}\left[\boldsymbol{\Delta}_{X},\boldsymbol{\Delta}_{Y}\right]=\sum_{i}r_{0,i}\Delta_{i}^{2}+\sum_{i,j}u_{ij}\Delta_{i}^{2}\Delta_{j}^{2},
\label{effective_S3}
\end{equation}
 with the coefficients:
\begin{eqnarray}
r_{0,i} & = & \frac{2}{u_{\mathrm{s}}}+\frac{1}{2}\int_{k}G_{\Gamma,k}G_{i,k},\nonumber \\
u_{ij} & = & \frac{1}{8 N}\int_{k}G_{\Gamma,k}^{2}G_{i,k}G_{j,k}.
\label{GL}
\end{eqnarray}
We used the notation $\int_{k}=T\sum \limits_{n}\int\frac{d^{d}k}{\left(2\pi\right)^{d}}$.
The index $k=\left(\mathbf{k},\omega_{n}\right)$ combines the momentum $\mathbf{k}$
and the Matsubara frequency $\omega_{n}=\left(2n+1\right)\pi T$.

Using the identities 
\begin{eqnarray}
\int_{k} G_{\Gamma,k}G_{X,k} & = & \int_{k} G_{\Gamma, k}G_{Y, k},\nonumber \\
\int_{k} G_{\Gamma,k}^{2}G_{X,k}^{2} & = & \int_{k} G_{\Gamma,k}^{2}G_{Y,k}^{2},
\end{eqnarray}
valid because the underlying Hamiltonian obeys the full C$_4$ symmetry, 
we can write the action in the more convenient form:
\begin{eqnarray}
S_{\mathrm{eff}}[\boldsymbol{\Delta}_{X},\boldsymbol{\Delta}_{Y}] & = & r_{0}({\mathbf{\Delta}}_{X}^{2}+{\mathbf{\Delta}}_{Y}^{2})+\frac{u}{2}({\mathbf{\Delta}}_{X}^{2}+{\mathbf{\Delta}}_{Y}^{2})^{2} \nonumber \\
 & & -\frac{g}{2}({\mathbf{\Delta}}_{X}^{2}-{\mathbf{\Delta}}_{Y}^{2})^{2},
 \label{action_ug1}
\end{eqnarray}
with coefficients
\begin{eqnarray}
r_{0} & = & \frac{2}{u_{s}}+\frac{1}{2}\int G_{X,k}G_{\Gamma,k},\nonumber \\
u & = &  \frac{1}{16 N}\int_{k}G_{\Gamma,k}^{2}(G_{X,k}+G_{Y,k})^{2},\nonumber \\
g & = & - \frac{1}{16 N}\int_{k}G_{\Gamma,k}^{2}(G_{X,k}-G_{Y,k})^{2}.
\label{action_ug}
\end{eqnarray}
 $r_{0}$, $u$ and $g$ have been calculated as a function of temperature and band parameters in Ref. \onlinecite{Fernandes12}. It was found that $u>0$  and $u>g$ in general. The coupling $g$ vanishes for circular electron pockets, but is positive for a non-zero ellipticity. 

\subsection{Nematic susceptibility in the large-$N$ expansion} 
In order to investigate the possibility of the nematic transition occuring before the magnetic transition, we follow the steps of Ref. \onlinecite{Fernandes12}, and introduce two auxiliarly Hubbard-Stratonovich scalar fields $\phi$ and $\psi$ to decouple the quartic terms in the action (\ref{action_ug1}); $\phi \rightarrow {\mathbf{\Delta}}_{X}^{2}-{\mathbf{\Delta}}_{Y}^{2}$ and $\psi \rightarrow {\mathbf{\Delta}}_{X}^{2}+{\mathbf{\Delta}}_{Y}^{2}$. The resulting effective action is given by
\begin{eqnarray}
S_{\mathrm{eff}}&=&\int_{q}\chi_{q}^{-1}\left(\boldsymbol{\Delta}_{X}^{2}+\boldsymbol{\Delta}_{Y}^{2}\right)+\int_{x}\left(\frac{\phi^{2}}{2g}-\frac{\psi^{2}}{2u}\right)\nonumber \\
 & &+\int_{x}\psi\left(\boldsymbol{\Delta}_{X}^{2}+\boldsymbol{\Delta}_{Y}^{2}\right)+\int_{x}\left ( \phi + h_{\mathrm{n}}\right )\left(\boldsymbol{\Delta}_{X}^{2}-\boldsymbol{\Delta}_{Y}^{2}\right),
\nonumber \\
\label{S_Delta_phi}
\end{eqnarray}
and we have added a field $ h_{\mathrm{n}}$ conjugate to the nematic order parameter $\Delta_{X}^{2}-\Delta_{Y}^{2}$. This term is needed in order to calculate the susceptibility of the nematic order parameter. A finite value of $\phi$ implies non-zero expectation value of $\frac{\phi}{g}=\langle {\mathbf{\Delta}}_{X}^{2}-{\mathbf{\Delta}}_{Y}^{2} \rangle \neq 0$ and the system develops nematic order. The large-$N$ mean field value of $\psi$ is always non-zero and describes the strength of magnetic fluctuations.
In case of split magnetic and structural phase transitions, there is no magnetic order right below the structural transition temperature, i.e $\langle {\mathbf{\Delta}}_{X,Y}\rangle=0$. Next we integrate out the $N^2-1$ component fields ${\mathbf{\Delta}}_{X,Y}$. 
If we further rescale the coupling constants to $\tilde g = g (N^2-1)$ and $\tilde u= u(N^2-1)$, required to reach a sensible large-$N$ limit, the effective action can be written as
\begin{eqnarray}
S_{\mathrm{eff}}[\psi,\phi]&=&N^2\int_{q}\left\{ \frac{\phi^{2}}{2\tilde g}-\frac{\psi^{2}}{2 \tilde u} \right \}
\nonumber \\
& &+\frac{N^2}{2} \int_{q}\left\{\log\left[\left(\chi_{q}^{-1}+ \psi  \right)^{2}-\left ( \phi+h_{\mathrm{n}}\right )^{2}\right] \right\}.
\nonumber \\
\label{S_phi_N}
\end{eqnarray}
We note that the effective action (\ref{S_phi_N}) has an overall pre-factor $N^2$. For $N \gg 1$  the integral over the fields $\phi$ and $\psi$ can be performed via the saddle-point method, i.e. by analyzing the extremum of the action. After solving for $\partial S_{\mathrm{eff}}\left[\phi,\psi\right]/\partial\phi=\partial S_{\mathrm{eff}}\left[\phi,\psi\right]/\partial\psi=0$, we obtain the equations of state for $\phi$ and $\psi$:
\begin{eqnarray}
\frac{\psi}{\tilde u} & = & \int_{q}\frac{\chi_{q}^{-1}+\psi}{\left(\chi_{q}^{-1}+\psi \right )^{2}-\left ( \phi+h_{\mathrm{n}}\right )^{2}},\nonumber \\
\frac{\phi}{\tilde g} & = & \int_{q}\frac{\phi +h_{\mathrm{n}}}{\left(\chi_{q}^{-1}+\psi \right)^{2}-\left ( \phi+h_{\mathrm{n}}\right )^{2}}\label{phi_psi}.
\end{eqnarray}
By differentiating the second equation in (\ref{phi_psi}) with respect to the conjugate field, we find that, for small $\phi$
\begin{eqnarray}
\left. \frac{\partial \phi}{\partial h_{\mathrm{n}}} \right|_{h_{\mathrm{n}}=0} &=&
\frac{\tilde g \int_{k} \chi^2_{k} }{1-\tilde g \int_{k} \chi^2_{k}},
\end{eqnarray}
where, from now on, we have shifted $\chi^{-1}_{k} \rightarrow \chi^{-1}_{k} +\psi$, which simply corresponds to the re-normalisation of the mass term due to fluctuations. Similarly to the result of Ref. \onlinecite{Yamase}, we find that the electronic contribution to the susceptibility of the nematic order parameter $ {\mathbf{\Delta}}_{X}^{2}-{\mathbf{\Delta}}_{Y}^{2}$ is given by
\begin{eqnarray}
\chi_{\textrm{nem}}^{\textrm{el}}=\frac{ \int_{k} \chi^2_{k} }{1-\tilde g \int_{k} \chi^2_{k}},
\label{nematic_susc_el}
\end{eqnarray}
where $\chi^{-1}_{q}$ is the inverse magnetic susceptibility, and 
\begin{equation}
\tilde g  =- \frac{N}{16}\int_{k}G_{\Gamma,k}^{2}(G_{X,k}-G_{Y,k})^{2}
\label{tildeg}
\end{equation}
is the nematic coupling constant of the theory.
In Ref. \onlinecite{Fernandes12} it was found that for the classical phase transition in $d=2$ and $u/g>2$ the nematic transition pre-empts the magnetic transition, i.e. the transition lines are split. Also, the nematic transition was found to be of second order. This is the regime we are interested in.
What we have calculated so far is the purely electronic contribution to the nematic susceptibility. One, however needs to include the effect of the lattice, as was pointed out in Ref. \onlinecite{Indranil,  Kontani1}.
The nemato-elastic coupling is given by the following Hamiltonian
\begin{eqnarray}
\mathcal{H}_{\textrm{nem}}=\gamma_{\textrm{el}}\int d r \phi(r) \left ( \partial_{x} u_x - \partial_{y} u_y \right ),
\end{eqnarray}
where $\gamma_{\textrm{el}}$ is the nemato-elastic coupling constant and ${\bf u}=(u_x, u_y)$ the phonon displacement field. The phonons renormalize the nematic coupling constant to a frequency and momentum dependent coupling
\begin{eqnarray}
\tilde g(q,\omega)=\tilde g+\gamma_{\textrm{el}}^2 \frac{q^2}{c_{\rm s}^{0}q^2-\omega^2},
\end{eqnarray}
where $c_{\rm s}^{0}$ is the elastic constant and $q$ the momentum along the soft directions. In particular, if one wants to determine the location of the nematic phase transition, which is dictated by the condition of divergent nematic susceptibility, one needs to look at the static limit of the coupling constant, i.e. the limit where $\omega$ is set to zero. This leads to $ g_{\textrm{stat}}=\tilde g +\frac{\gamma_{\textrm{el}}^2}{c_{\rm s}^{0}}$.
The full nematic susceptibility, including the effect of the coupling to the lattice, in the large $N$ expansion is therefore given by 
\begin{eqnarray}
\chi_{\textrm{nem}}=\frac{ \int_{k} \chi^2_{k} }{1-g_{\textrm{stat}} \int_{k} \chi^2_{k}},
\label{nematic_susc}
\end{eqnarray}
where 
\begin{eqnarray}
g_{\textrm{stat}}=\tilde g + \frac{\gamma_{\textrm{el}}^2}{c_{\rm s}^{0}}.
\label{gstat}
\end{eqnarray}
\section{Raman response function}
\label{Raman}

Raman scattering is a valuable tool to study strongly correlated electronic systems \cite{Rudireview}, since it probes lattice, spin and electronic degrees of freedom. It has been used to extract information about the momentum structure and symmetry of the excitations in the cuprates \cite{Caprara11, Caprara09, Tassini, Caprara05} and pnictides. 
The differential photon scattering cross section in Raman spectroscopy is directly proportional to the structure factor $S$:
\begin{eqnarray}
S_{q}&=&-\frac{1}{\pi}
\left[1+n(\omega)\right]{\rm Im} R_q,
\label{sf}
\end{eqnarray}
which is related to the imaginary part to the Raman response function $R$
through the fluctuation-dissipation theorem. \cite{Devereaux} Here, $n(\omega)$ is the Bose-Einstein distribution function, and $q=({\bf q},\omega)$. Since the momentum of light is much smaller than the typical lattice momentum, one normally replaces ${\bf q} \approx 0$ in Eq. (\ref{sf}).

The Raman response functions measures correlations between ``effective charge density'' fluctuations $\tilde \rho$,
\begin{equation}
R( \omega)=\int_{0}^{1/T} d\tau\, e^{-i\omega\tau}
\langle \tilde\rho(\tau)\tilde\rho(0)\rangle.
\label{R_densities}
\end{equation}  
The effective density, weighted by the form factors that can be changed via the geometry of the photon polarization, is defined as
\begin{equation}
\tilde \rho_{k} = \sum_{i,k',\sigma} \gamma_{{\mathbf{k'}}} c^{\dagger}_{i,k+k',\sigma} c_{i,k',\sigma}.
\end{equation}
$\sigma$ is the spin index, $i$ the band index, and the operator $c^{\dagger}_{i,k,\sigma}$ creates an electron with spin $\sigma$ and momentum ${\bf k}$ in band $i$, where $i=X,Y,\Gamma$. The function $\gamma_{{\mathbf{k}}}$ is related to the incident and scattered photon polarization vectors and depends on the curvature of the bands. \cite{Devereaux} Here, we will consider the intra-orbital contributions to $\gamma_k$ only, as this is the dominant process. The multi-orbital nature of different bands has been pointed out in Ref. \onlinecite{Valenzuela}.

In order to determine the Raman response function, we couple an external source field to the weighted densities and  introduce the generating functional $W_{h}$ according to:
\begin{eqnarray}
W_{h} &=& \frac{1}{Z} \int d\Delta_{i}d\Psi\mathrm{e}^{-S\left[\Psi,\Delta_{i}\right]-\Psi^\dagger \mathcal{V}_{h} \Psi},
\nonumber \\
Z &=& \int d\Delta_{i}d\Psi\mathrm{e}^{-S\left[\Psi,\Delta_{i}\right]},
\label{Wh}
\end{eqnarray}
where $S\left[\Psi,\Delta_{i}\right]$ is given in Eq. (\ref{S}). The elements of the matrix $ \mathcal{V}_{h}$ in momentum/frequency, spin and band space are
\begin{eqnarray}
\mathcal{V}_{h,k_{1} k_{2} \sigma \sigma' i j}=h_{k_{1}-k_{2}} \gamma_{{\bf k_1}}\delta_{\sigma \sigma'}\delta_{ij},
\end{eqnarray}
with $h$ being the field conjugate to the effective density. The Raman response function (\ref{R_densities}) is  obtained by differentiating
the generating functional $W_{h}$ (\ref{Wh}) with respect to the conjugate field $h$:
\begin{eqnarray}
R_{q} & = &\left. \frac{\delta ^2 W_{h}}{\delta h_{q} \delta h_{-q}} \right|_{h=0}.
\label{Rh}
\end{eqnarray}
Due to the single particle character of the source term, the generating functional Eq. (\ref{Wh}) can be written in the form
\begin{eqnarray}
W_{h} & = &\frac{1}{Z} \int d\Delta_{i}d\Psi \mathrm{e}^{\int \Psi^\dagger \mathcal{G}^{-1}_{\Delta, h} \Psi -
\frac{2}{u_{\mathrm{s}}} \int_{x}\left(\Delta_{X}^{2}+\Delta_{Y}^{2}\right)}
\nonumber \\
\mathcal{G}_{\Delta, h}^{-1} & = & \mathcal{G}_{0}^{-1}-\mathcal{V}_{\Delta}-\mathcal{V}_{h}.
\end{eqnarray}
Since $W_{h}$ contains the action that is quadratic in fermions, we integrate out the fermions and obtain:
\begin{eqnarray}
W_{h} & = & \frac{1}{Z} \int d\Delta_{i} \mathrm{e}^{- S_{h}\left [ \Delta_{i}\right ]},
\nonumber \\
S_{h} \left [ \Delta_{i}\right ] & = & \frac{2}{u_{\mathrm{s}}} \int_{x}\left(\boldsymbol{\Delta}_{X}^{2}+\boldsymbol{\Delta}_{Y}^{2}\right)-\mathrm{Tr}\ln \left ( \mathcal{G}^{-1}_{\Delta, h}\right ).
\label{G_Delta_h}
\end{eqnarray}
We further expand:
\begin{equation}
\mathrm{Tr}\ln \left ( \mathcal{G}^{-1}_{\Delta, h}\right ) =   \mathrm{Tr}\ln \left ( \mathcal{G}^{-1}_{\Delta}\right ) - \sum_{n=1}^{\infty}\frac{ \mathrm{Tr} \left ( \mathcal{G}_{\Delta} \mathcal{V}_h\right )^n }{n}.
\end{equation}
Then, using (\ref{G_Delta_h}) and (\ref{Rh}), 
\begin{eqnarray}
R_{q} &=&\frac{1}{Z} \int d \Delta_i e^{-S_{\mathrm{eff}}[\Delta_X, \Delta_Y]}
\nonumber \\
& &\frac{\delta^2}{\delta h_{q} \delta h_{-q}} \left. \exp{ \left [-\mathrm{Tr}\left ( \mathcal{G}_{\Delta}\mathcal{V}_{h}\right )-\frac{1}{2}\mathrm{Tr}\left (\mathcal{G}_{\Delta}\mathcal{V}_{h}\right )^2\right ]} \right|_{h=0}.
\nonumber \\
\label{RGamma}
\end{eqnarray}
Here $S_{\mathrm{eff}}[\Delta_X,\Delta_Y] =S_{h}[\Delta_i]\vert_{h=0}$ is the effective action given by
(\ref{effective_S2}).
We define the matrix
\begin{eqnarray}
\Gamma^{q} &=& \frac{{\delta \mathcal{V}_{h}}}{\delta h_{q}}.
\end{eqnarray}
\subsection{Self-energy and vertex correction diagrams}
\begin{figure}
\begin{centering}
\includegraphics[width=0.7\columnwidth]{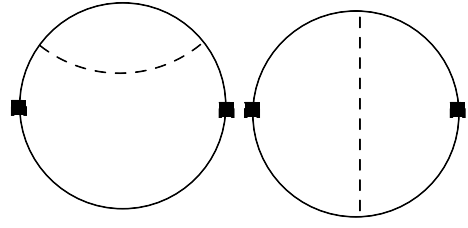} 
\par\end{centering}
\caption{Left: Contribution to the Raman response function that contains the self-energy correction to the fermionic propagator. Right: A diagram that contains a vertex renormalization correction.}
\label{selfenergy}
\end{figure}

Next, we analyse the leading order contributions to the Raman response function. These arise from the self-energy and vertex correction diagrams depicted in Fig. \ref{selfenergy}. 
Both of these diagrams arise from differentiating the second term in the exponential (\ref{RGamma}) twice with respect to $h$
\begin{eqnarray}
R_{q} ^{V,S}&=&\frac{1}{Z} \int d \Delta_i e^{-S_{\mathrm{eff}}[\Delta_X,\Delta_Y]}  \mathrm{Tr}\left [ \left ( \mathcal{G}_{\Delta }\Gamma \right )^2 \right ],
\label{VS}
\end{eqnarray}
and we replace $S_{\mathrm{eff}} \rightarrow S_0$, where $S_0$ is the quadratic action given by
\begin{eqnarray}
S_0 \left [ \Delta_i \right ]=\frac{2}{u_{\mathrm{s}}}\int \left(\Delta_{X}^{2}+\Delta_{Y}^{2}\right) + \frac{1}{2} \mathrm{Tr}\left (\mathcal{G}_{0}\mathcal{V}_{\Delta}\right )^2.
\label{quadraticaction}
\end{eqnarray}
In order to get the vertex correction, we replace both $\mathcal{G}_{\Delta}$ in (\ref{VS}) by $\mathcal{G}_{\Delta} \rightarrow\mathcal{G}_{0} \mathcal{V}_{\Delta} \mathcal{G}_{0}$, which comes from the perturbative expansion of Eq. (\ref{invGdelta}):
\begin{eqnarray}
R_{q}^{V} & = &\frac{1}{Z} \int d \Delta_i e^{-S_{0}[\Delta_X,\Delta_Y]}  \mathrm{Tr}\left [ \left ( \mathcal{G}_{0} \mathcal{V}_{\Delta} \mathcal{G}_{0} \Gamma \right )^2 \right ].
\label{V}
\end{eqnarray}
In order to get the self-energy correction, we replace one of $\mathcal{G}_{\Delta}$ in (\ref{VS}) by $\mathcal{G}_{\Delta} \rightarrow\mathcal{G}_{0}$, and the other one
by $\mathcal{G}_{\Delta} \rightarrow \left ( \mathcal{G}_{0} \mathcal{V}_{\Delta} \right )^2 \mathcal{G}_{0}$ to get
\begin{eqnarray}
R_{q}^{S} & = & \frac{2}{Z} \int d \Delta_i e^{-S_{0}[\Delta_X,\Delta_Y]}  \mathrm{Tr}\left [ \mathcal{G}_{0} \Gamma \left ( \mathcal{G}_{0} \mathcal{V}_{\Delta}  \right )^2  \mathcal{G}_{0} \Gamma \right ].
\nonumber \\
\label{V}
\end{eqnarray}
Due to the integral over the square of the $\gamma_{{\mathbf{k}}}$ factor, the self-energy and vertex corrections occur in all symmetry channels. If one evaluates the sum $R^{S}+R^{V}$ explicitly, in the hot-spot approximation, one finds that there are partial cancellation in the $A_{1g}$ and in the $B_{1g}$ channels, and no cancellations in the $B_{2g}$ channel. One can easily show that in $d=2$
\begin{eqnarray}
R^{S}+R^{V} \propto \int_{{\bf q}}\frac{1}{r_0+q^2} \propto \log{\xi},
\end{eqnarray}
where we have used $r_0 =\xi^{-2}$, where $\xi$ is the magnetic correlation length. 
\subsection{Leading order Aslamazov-Larkin diagrams}
\begin{figure}
\begin{centering}
\includegraphics[width=1\columnwidth]{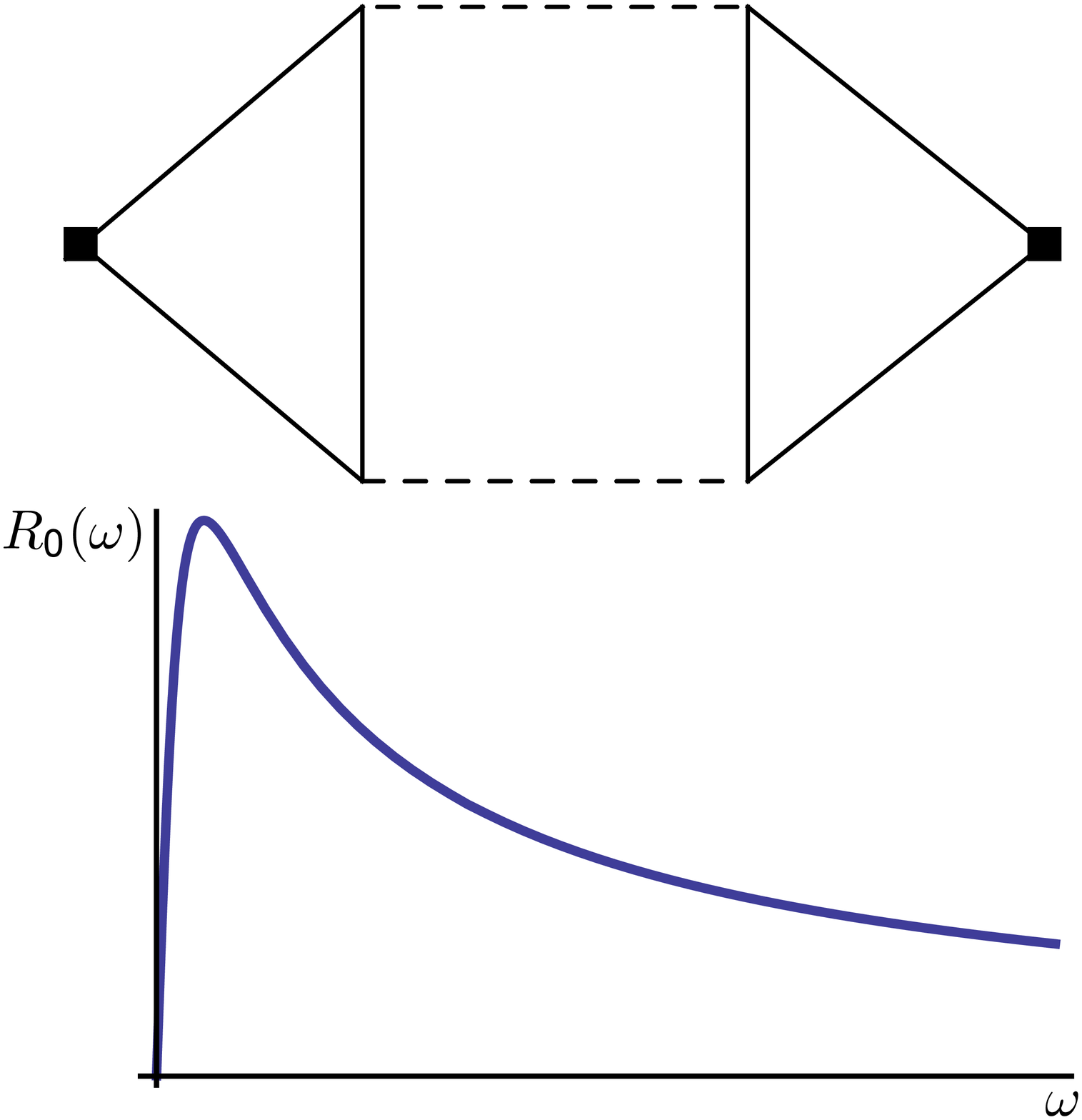} 
\par\end{centering}
\caption{Leading order Aslamazov-Larkin diagram. Raman vertices are denoted by black squares. Imaginary part of the Raman response function as a function of frequency $\mathrm{Im} R_0(\omega)$, in $d=2$.}
\label{leading}
\end{figure}
The Aslamazov-Larkin contribution to the Raman response function, analyzed in Ref. \onlinecite{Caprara05} arises from differentiating the first term inside the exponential in (\ref{RGamma}) twice, and from replacing $\mathcal{G}_{\Delta} \rightarrow \left ( \mathcal{G}_{0} \mathcal{V}_{\Delta} \right )^2 \mathcal{G}_{0}$, which comes from the perturbative expansion of Eq. (\ref{invGdelta}):
\begin{eqnarray}
R_{q} & = &\frac{1}{Z} \int d \Delta_i e^{-S_{\mathrm{eff}}[\Delta_X,\Delta_Y]} \left [ \mathrm{Tr}\left ( \left ( \mathcal{G}_{0} \mathcal{V}_{\Delta} \right )^2 \mathcal{G}_{0}\Gamma \right ) \right ]^{2}.
\label{diagramatics_AL}
\nonumber \\
\end{eqnarray}
Here $S_{\mathrm{eff}}[\Delta_X,\Delta_Y] =S_{h}[\Delta_i]\vert_{h=0}$ is the effective action given by
(\ref{effective_S2}).

As we will see below, the key assumption of a description based on the Aslamasov-Larkin diagrams is that one neglects the interactions between spin fluctuations. In other words, one approximates the effective action in
(\ref{diagramatics_AL}) by quadratic action Eq. (\ref{quadraticaction}).
While this assumption is frequently justified, it is not allowed in the theory of spin-driven nematicity, as we will show later.

The leading order Aslamazov-Larkin diagram, depicted in Fig. \ref{leading} can be calculated as
\begin{eqnarray}
R_{0}(\omega) &=&T \sum_{i=X,Y,n}\int_{{\mathbf{q}}}  \Lambda^{2}_{i}({\mathbf{q}},\Omega_n,\omega) \chi ({\mathbf{q}},\Omega_{n}) \chi ({\mathbf{q}}, \Omega_{n}-\omega)
\nonumber \\
\label{leading_AL}
\end{eqnarray}
with
\begin{eqnarray}
\Lambda_{i}({\mathbf{q}},\Omega,\omega) &=& \Lambda_{i}^{(1)}({\mathbf{q}},\Omega,\omega)+\Lambda_{i}^{(2)}(-{\mathbf{q}},-\Omega,-\omega),
\nonumber \\
\Lambda_{i}^{(1)}({\mathbf{q}},\Omega,\omega) &=&T\sum \limits_{n} \int_{{\mathbf{k}}}
 \gamma_{{\mathbf{k}}}G_{\Gamma}({\mathbf{k}},\nu_{n}-\omega)G_{\Gamma}({\mathbf{k}},\nu_{n})
 \nonumber \\
& &\times G_{i}({\mathbf{k}}-{\mathbf{q}}, \nu_{n}- \Omega ),
\nonumber \\
\Lambda_{i}^{(2)}({\mathbf{q}},\Omega,\omega)&=&T\sum \limits_{n} \int_{{\mathbf{k}}}
 \gamma_{{\mathbf{k}}}G_{i}({\mathbf{k}},\nu_{n}-\omega)G_{i}({\mathbf{k}},\nu_{n})
 \nonumber \\
& &\times G_{\Gamma}({\mathbf{k}}-{\mathbf{q}}, \nu_{n}- \Omega ),
\label{lambda}
\end{eqnarray}
similar to what was found in Ref. \onlinecite{Indranil2}.

\subsubsection{Raman response in different symmetry channels}
In the concept of the pairing symmetry in high-temperature superconductors successful theoretical models supported by experiments have been developed in order to explain the symmetry sensitivity of the Raman response function. \cite{Devereauxdxy} Similarly, here, before we explicitly evaluate the leading order Aslamazov-Larkin diagram, we analyze the contribution to it in the various symmetry channels. Higher order corrections that will be discussed later do not alter this symmetry-based analysis. We show that the Aslamazov-Larkin diagram, given by Eq. (\ref{leading_AL}) and (\ref{lambda}), only supports the $B_{1g}$ and the $A_{1g}$ symmetry channels.
Let us consider the structure of the terms in (\ref{leading_AL}) which arise from
\begin{eqnarray}
R_0^{(11)}(\omega) \coloneqq & &T \sum_{i=X,Y,n}\int_{{\mathbf{q}}}  \left [ \Lambda^{(1)}_{i}({\mathbf{q}},\Omega_n,\omega)\right ]^2 \nonumber \\
& & \times \chi ({\mathbf{q}},\Omega_{n})
\chi ({\mathbf{q}}, \Omega_{n}-\omega).
\label{r11}
\end{eqnarray}
The term (\ref{r11}) can be re-written in the following form
\begin{eqnarray}
R_0^{(11)}(\omega)=& & \frac{T}{2} \sum_{n}\int_{{\mathbf{q}}} \int_{{\mathbf{k}}} \int_{{\mathbf{p}}} 
 \gamma_{{\mathbf{k}}}  \gamma_{{\mathbf{p}}} \chi ({\mathbf{q}},\Omega_n)
\chi ({\mathbf{q}}, \Omega_n-\omega)\nonumber \\
& &\times (  E_{A_{1g}}(\omega,\Omega_n,{\mathbf{k}},{\mathbf{q}})E_{A_{1g}}(\omega,\Omega_n,{\mathbf{p}},{\mathbf{q}})
\nonumber \\
& & +E_{B_{1g}}(\omega,\Omega_n,{\mathbf{k}},{\mathbf{q}})E_{B_{1g}}(\omega,\Omega_n,{\mathbf{p}},{\mathbf{q}})
 ),\nonumber \\
\end{eqnarray}
where we have classified the appropriate combinations of Green's functions according to their symmetry into
\begin{eqnarray}
E_{A_{1g}}(\omega,\Omega_n,{\mathbf{k}},{\mathbf{q}})=&& T \sum_{m}
G_{\Gamma}({\mathbf{k}},\nu_{m}-\omega)G_{\Gamma}({\mathbf{k}},\nu_{m})
\nonumber \\
& & \times G^{(+)}({\mathbf{k}}-{\mathbf{q}},\nu_{m}-\Omega_n),
\nonumber \\
E_{B_{1g}}(\omega,\Omega_n,{\mathbf{k}},{\mathbf{q}})=&&T \sum_{m}
G_{\Gamma}({\mathbf{k}},\nu_{m}-\omega)G_{\Gamma}({\mathbf{k}},\nu_{m})
\nonumber \\
& & \times G^{(-)}({\mathbf{k}}-{\mathbf{q}},\nu_{m}-\Omega_n),
\nonumber \\
\label{e}
\end{eqnarray}
and we have defined $G^{(\pm)}=G_X \pm G_Y$. From Eq. (\ref{e}), we see that the response will be non-zero only for $\gamma$ factors in the $A_{1g}$ or the $B_{1g}$ symmetry. Similarly, using the same line of arguments, one can show that all other terms in (\ref{leading_AL}) support the $A_{1g}$ or the $B_{1g}$ symmetry only. We have thus ruled out the response in the $B_{2g}$ channel.
\subsubsection{ Explicit calculation of the leading order Aslamazov-Larkin diagram}
\label{leading_calculation}
The leading order Aslamazov-Larkin diagram has been evaluated in Ref. \onlinecite{Caprara05}, assuming that the main contribution comes from the hot-spot regions and that the momenta of the fluctuations are peaked around ${\mathbf{q}} \approx {\mathbf{Q}}_{X,Y}$. After the analytic continuation to the real frequencies, we found that the imaginary part of the Raman response function, which is a quantity of experimental interest, is given by
\begin{eqnarray}
\mathrm{Im} R_{0}(\omega +i0^{+}) &=&\int_{-\infty}^{\infty}\frac{{\textrm{d}}\epsilon}{\pi} \left [ n(\epsilon)-n(\epsilon+\omega) \right ]
\nonumber \\
& &\times \int_{{\mathbf{q}}}\mathrm{Im}\left [ \chi^{R}(\epsilon,{\mathbf{q}})\right ]\mathrm{Im}\left [ \chi^{R}(\epsilon+\omega,{\mathbf{q}}) \right ],
\nonumber \\
\label{R0_integral}
\end{eqnarray}
with the spin propagator in the tetragonal phase given by:
\begin{eqnarray}
\chi^{R}({\mathbf{q}},\Omega) & = & \frac{1}{r_0+{\mathbf{q}}^{2}-i\Omega},
\end{eqnarray} 
where $r_0$ tunes the distance from the magnetic transition, see Eq. (\ref{action_ug}).
In $d=2$ the ${\mathbf{q}}$ integral in Eq. (\ref{R0_integral}) in can be performed exactly, which
leads to the following expression
\begin{eqnarray}
\mathrm{Im} \left [ R_{0}(\omega +i0^{+})\right ]_{d=2}
 =& & \int_{0}^{\infty}{\textrm{d}}\epsilon \left [ n(\epsilon_{+})-n(\epsilon_{-}) \right ]\frac{\epsilon_{+}\epsilon_{-}}{\epsilon_{+}^{2}-\epsilon_{-}^{2}}
 \nonumber \\
& & \times \left [ F(\epsilon_{+})-F(\epsilon_{-}) \right ],
\label{2draman}
\end{eqnarray}
with 
\begin{equation}
F(x)  = \frac{1}{x}\left(\arctan{\frac{r_0}{x}}-\frac{\pi}{2}sgn(x)\right).
\end{equation}
We defined $\epsilon_{\pm}=\epsilon\pm\omega/2$. The plot of the function (\ref{2draman}) is shown in Fig. {\ref{leading}}.
In particular one can deduce that, in the regime where temperature $T$ is the biggest scale, $T \gg r_0$, $R_{0}(\omega)_{d=2}\simeq\frac{\omega T}{r_0^{2}}$ for small frequencies $\omega$, while the amplitude of the Raman response function scales as $R^{\textrm{max}}_{0}(\omega)_{d=2}\simeq\frac{T}{r_0}$ in this regime. 

In summary, we have shown that the leading order Aslamazov-Larkin diagram gives a non-zero response in the $B_{1g}$ and $A_{1g}$ symmetries only. It predicts the divergence of the Raman response at the magnetic transition, and does not carry any signatures of the structural transition. We therefore need to go beyond the leading order Aslamazov-Larkin diagram.

\subsection{Higher order Aslamazov-Larkin-like diagrams}

Next, we go beyond the quadratic action approximation for $S_{\mathrm{eff}}$ in (\ref{diagramatics_AL}), and include the full quartic action to evaluate the Raman response function. As we will show, diagrammatically this corresponds to inserting a series of fermionic boxes that resemble the structure of the nematic coupling constant $\tilde g$ into the leading order Aslamazov-Larkin diagram in the $B_{1g}$ symmetry. These diagrams take into account the collisions between spin fluctuations which were not accounted for in the leading order Aslamazov-Larkin diagram. 

\begin{figure}
\begin{centering}
\includegraphics[width=1\columnwidth]{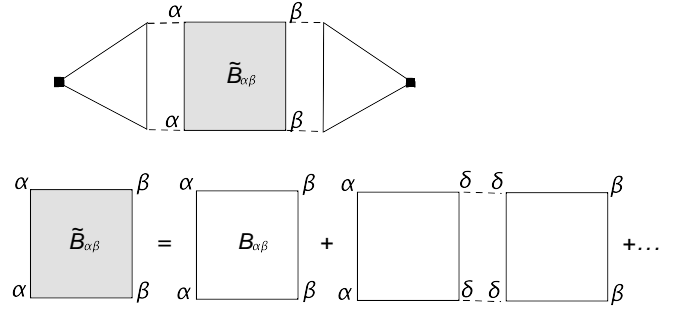} 
\par\end{centering}
\caption{Re-summed Raman response function. The resummed box $\tilde{B}_{\alpha\beta}$
is shaded grey. The first index of the matrix $B$ denotes the type
$\alpha=X,Y$ of entering spin fluctuations, and the second index
the type of exiting spin fluctuations. We insert the grey shaded box
into the Raman reponse, and make some further simplifications to evaluate
the Raman response function (see the main text).}
\label{resum}
\end{figure}

First we show how these terms arise from the diagrammatic expansion. We start from Eq. (\ref{diagramatics_AL}), but this time we go beyond the quadratic approximation for the effective action, and include quartic terms
\begin{eqnarray}
R_{q} & = &\frac{1}{Z} \int d \Delta_i e^{-S_{\mathrm{eff}}[\Delta_{i}]} \left [ \mathrm{Tr}\left ( \left ( \mathcal{G}_{0} \mathcal{V}_{\Delta} \right )^2 \mathcal{G}_{0}\Gamma \right ) \right ]^{2}
\label{S0}
\end{eqnarray}
where
\begin{eqnarray}
S_{\mathrm{eff}} \left[\boldsymbol{\Delta}_{i} \right ]
& = & S_{0}\left[\boldsymbol{\Delta}_{i} \right ]+
\frac{1}{4}\mathrm{Tr} \left (\mathcal{G}_{0}\mathcal{V}_{\Delta} \right )^{4},
\end{eqnarray}
with the bare action
\begin{eqnarray}
S_0\left[\boldsymbol{\Delta}_{i} \right ] & = & \frac{2}{u_{\mathrm{s}}}\int_{x}\left(\boldsymbol{\Delta}_{X}^{2}+\boldsymbol{\Delta}_{Y}^{2}\right)+\frac{1}{2}\mathrm{Tr} \left (\mathcal{G}_{0}\mathcal{V}_{\Delta} \right )^{2}.
\end{eqnarray}
We further expand the exponential 
\begin{equation}
e^{-\frac{1}{4}\mathrm{Tr} \left (G_{0}\mathcal{V}_{\Delta} \right )^{4}} \approx
\sum\limits_{m=0}^{\infty} \frac{1}{m!} \left [ \frac{-1}{4}\mathrm{Tr} \left (\mathcal{G}_{0}\mathcal{V}_{\Delta} \right )^{4} \right ]^{m}
\end{equation}
to obtain
\begin{equation}
R_{q}  = \sum_{m=0}^{\infty}\frac{1}{m!} R_{q}^{(m)},
\label{R_renormalized}
\end{equation}
where we averaged the following terms with respect to the Gaussian collective spin action:
\begin{equation}
R_{q}^{(m)} = \left  \langle\left [ \frac{-1}{4}\mathrm{Tr}\left ( \mathcal{G}_{0}\mathcal{V}_{\Delta} \right )^4 \right ]^m 
\left [ \mathrm{Tr}\left ( \left ( \mathcal{G}_{0} \mathcal{V}_{\Delta} \right )^2 \mathcal{G}_{0}\Gamma \right ) \right ]^{2}
\right \rangle_{S_0}.
\label{R_renormalized}
\nonumber \\
\end{equation}
In order to evaluate the expectation values one performs contractions of the $\Delta$ fields. We obtain a series of diagrams that look like the leading order Aslamazov-Larkin diagram with an arbitrary number of inserted fermionic boxes, depicted in Fig. \ref{resum}. 

The higher order diagrams effectively take collisions between spin fluctuations into account, which have been neglected in the leading order Aslamazov-Larkin diagram. As one approaches the transition line, collisions between spin fluctuations become more and more important and one would anticipate significant changes in the Raman response function due to these processes. As we will show, the re-summation of boxed Aslamazov-Larkin diagrams will lead to the maximum of the Raman response function at the structural phase transition in the $B_{1g}$ channel, and the suppression of the response in the $A_{1g}$ channel.

The next task is to re-sum an infinite number of such diagrams. Every box can be characterized by two indices: the first one denotes the type of incoming spin fluctuations, this can be either $X$ or $Y$ and the second one denotes the type of exiting spin fluctuation. Let us denote this box $B_{\alpha \beta}$.
Summing all boxed diagrams can be most efficiently expressed as:
\begin{eqnarray}
{R}(\omega) &=& R_{0}(\omega)+ T^2 \sum \limits_{\Omega,\Omega'}  \int_{{\mathbf{q}},{\mathbf{q}}'} \Lambda_{\alpha}(\omega,\Omega,{\mathbf{q}}) \nonumber \\
&\times & \chi ({\mathbf{q}},\Omega)\chi ({\mathbf{q}},\Omega-\omega) \nonumber \\
& \times & \tilde{B}_{\alpha\beta}({\mathbf{q}},{\mathbf{q}}',\Omega,\Omega',\omega)\chi ({\mathbf{q}}',\Omega') 
\nonumber \\
& \times & \chi ({\mathbf{q}}',\Omega' -\omega)
 \Lambda_{\beta}(\omega,\Omega',{\mathbf{q}}').
\end{eqnarray}
For our analysis it is sufficient to calculate the box $B_{\alpha \beta}$ at momenta ${\mathbf{q}},{\mathbf{q'}} \approx {\mathbf{Q_{X,Y}}}$ and zero frequencies, which is justified for small incoming Raman frequency $\omega$, and in the classical regime relevant near a finite temperature phase transition. We write the Raman reponse function in the tetragonal phase:
\begin{eqnarray}
{R}(\omega) &\approx &  R_{0}(\omega)+ \int_{{\mathbf{q}},{\mathbf{q}}'} \Lambda_{\alpha}(\omega,0,{\mathbf{q}})\tilde{B}_{\alpha\beta}
\nonumber \\
& &\times \chi^{2}({\mathbf{q}},0)\chi^{2}({\mathbf{q}}',0) \Lambda_{\beta}(\omega,0,{\mathbf{q}}'),
\end{eqnarray}
where $R_{0}(\omega)$ is the leading order diagram. 

The symmetry of the fermionic triangle is such that
\begin{eqnarray}
\Lambda^{B_{1g}}_{X} & = & -\Lambda^{B_{1g}}_{Y},
\nonumber \\
\Lambda^{A_{1g}}_{X} & = & \Lambda^{A_{1g}}_{Y}.
\end{eqnarray}
This relation can be obtained by simply performing a coordinate system
rotation by $\pi/2$ inside the momenta integrals in (\ref{lambda}). This allows us to explicitly perform the matrix multiplication, which yields:
\begin{eqnarray}
\tilde{R}_{B_{1g}}(\omega) & = & 
 R_{0}(\omega)+R_{0}(\omega)(\tilde{B}_{XX}-\tilde{B}_{XY})\int_{{\mathbf{q}}}\chi^{2}({\mathbf{q}},0),
 \nonumber \\
 \tilde{R}_{A_{1g}}(\omega) & = & 
 R_{0}(\omega)+R_{0}(\omega)(\tilde{B}_{XX}+\tilde{B}_{XY})\int_{{\mathbf{q}}}\chi^{2}({\mathbf{q}},0).
 \nonumber \\
\end{eqnarray}

Next we need to determine an expression for the full box $\tilde{B}_{\alpha \beta}$, i.e. perform a sum over the leading box-diagrams within the $1/N$ expansion. This is illustrated in Fig. \ref{resum} and can be written as:
\begin{eqnarray}
\tilde{B}_{\alpha\beta} & = & B_{\alpha\beta}+B_{\alpha\delta}B_{\delta\beta}\int_{{\mathbf{q}}'}\chi^{2}({\mathbf{q}}',0)+...\nonumber \\
 & = & \sum_{m=1}^{\infty}(B^{m})_{\alpha\beta}\left (\int_{{\mathbf{q}}}\chi^{2}({\mathbf{q}},0) \right )^{m-1},
\end{eqnarray}
The matrix $B$ was deduced from Eq. (\ref{R_renormalized}) and Eq. (\ref{tm}). For details about explicit evaluation of the $SU(N)$ trace pre-factor (which arises from contractions of products of ${\mathbf{\lambda}}$ matrices in (\ref{V_delta})) for boxed diagram containing arbitrary number of boxes $m$, please see Appendix \ref{trace_prefactor}. The matrix $B$ of irreducible boxes is then given as
\begin{eqnarray}
B & =-\frac{N}{8} & \left(\begin{matrix}g_{XX} & g_{XY}\\
g_{XY} & g_{XX}
\end{matrix}\right ) 
\label{bmatrix}
\end{eqnarray}
where we used the abbreviation 
\begin{eqnarray}
g_{XX} & = & \int_{k}G_{\Gamma, k}^{2}G_{X,k}^{2},\nonumber \\
g_{XY} & = & \int_{k}G_{\Gamma, k}^{2}G_{X,k}G_{Y,k},
\end{eqnarray}
and  used that  by symmetry: $\int_{k}G_{\Gamma,k}^{2}G_{X,k}^{2}=\int_{k}G_{\Gamma,k}^{2}G_{Y,k}^{2}$. 

The $m$th power of the matrix $B$ is given by
\begin{eqnarray}
B^{m} & = &\frac{1}{2}\left ( \frac{-N}{8} \right )^{m}\left(\begin{matrix} \left ( g_{+}^{m}+g_{-}^{m} \right ) &\left ( g_{+}^{m}-g_{-}^{m} \right ) \\
\left ( g_{+}^{m}-g_{-}^{m} \right )& \left ( g_{+}^{m}+g_{-}^{m} \right )
\end{matrix}\right),
\label{bmmatrix}
\end{eqnarray}
where $g_{\pm}=g_{XX} \pm g_{XY}$. From this analysis follows that 
\begin{eqnarray}
\tilde{R}_{B_{1g}}(\omega) & = & R_{0}(\omega)\sum_{m=0}^{\infty}\left(\frac{-N g_{-}}{8}\right)^{m}\left (\int_{{\mathbf{q}}}\chi^{2}({\mathbf{q}},0)\right )^{m}\nonumber \\
  & = & R_{0}(\omega)+R_{0}(\omega)\frac{\tilde g \int_{{\mathbf{q}}}\chi^{2}({\mathbf{q}},0)}{1-\tilde g\int_{{\mathbf{q}}}\chi^{2}({\mathbf{q}},0)},
  \nonumber \\
  \tilde{R}_{A_{1g}}(\omega) & = & R_{0}(\omega)\sum_{m=0}^{\infty}\left(\frac{-N g_{+}}{8}\right)^{m}\left (\int_{{\mathbf{q}}}\chi^{2}({\mathbf{q}},0)\right )^{m}\nonumber \\
  & = & R_{0}(\omega)+R_{0}(\omega)\frac{\tilde u \int_{{\mathbf{q}}}\chi^{2}({\mathbf{q}},0)}{1+\tilde u\int_{{\mathbf{q}}}\chi^{2}({\mathbf{q}},0)},
\label{raman_final}
\end{eqnarray}
where 
\begin{equation}
\tilde g  = -\frac{N}{16}\int_{k} G_{\Gamma,k}^{2}(G_{X,k}-G_{Y,k})^{2}
\end{equation}
 is precisely the nematic coupling constant of Eq. (\ref{action_ug}) for the effective action, and $\tilde u$ is the other quartic term in Eq. (\ref{action_ug}), with $\tilde u>0$, as found in Ref \onlinecite{Fernandes12}. From (\ref{raman_final}), we see that the Raman response in the $A_{1g}$ channel gets suppressed, due to the term in the denominator, which grows as one approaches the transition. On the other hand, in the $B_{1g}$ channel, after performing the analytic continuation to real frequencies and taking the imaginary part, we get that
\begin{eqnarray}
{\rm Im}\tilde{R}_{B_{1g}}(\omega) & = & {\rm Im} \left [R_{0}(\omega)\right ] \left ( 1+ \tilde g {\chi}_{\mathrm{nem}}^{\mathrm{el}} \right),
\label{semifinal}
\end{eqnarray}
where 
\begin{eqnarray}
{\chi}_{\mathrm{nem}}^{\mathrm{el}}=\frac{ \int_{{\mathbf{q}}}\chi^{2}({\mathbf{q}},0)}{1-\tilde g\int_{{\mathbf{q}}}\chi^{2}({\mathbf{q}},0)}
\label{ramanchi}
\end{eqnarray}
is the electronic contribution to the nematic susceptibility calculated in the large $N$ limit \cite{Fernandes2010} for the model described in Sec. \ref{eff action}. As was pointed out in Ref. \onlinecite{Indranil}, the enhancement of the static nematic coupling constant (\ref{gstat}) does not enter the Raman response, due to the fact that the Raman response operates in the dynamical limit (${\mathbf{q}}=0$ and finite $\omega$), and the static and dynamic limits do not commute \cite{Indranil}. At the nematic / structural phase transition the nematic susceptibility (\ref{nematic_susc}) diverges, and
\begin{equation}
\left ( \tilde g +\frac{\gamma_{\rm el}^2}{c_{\rm s}^0}\right)\int_q \chi_{q}^2 =1.
\end{equation}
Consequently the Raman response function in the $B_{1g}$ channel, given by (\ref{semifinal}), has a maximum rather than a divergence at the structural phase transition.
\section{Conclusion}
\label{conclusion}
 In summary, we have shown that the Raman scattering can be used as a tool to probe the nematic phase in pnictides. We have presented a calculation that demonstrates that, in the low-frequency limit, and large $N$ limit, the Raman response function shows a clear maximum at the structural transition temperature in the $B_{1g}$ channel. 

In our model, the electronic nematic phase in pnictides is stabilized by spin-fluctuations associated with the striped phase, and occurs as a thin sliver above the magnetic transition temperature. In order to calculate the Raman response function, we have gone beyond the leading order Aslamazov-Larkin diagram, and included higher order diagrams that contain a series of quartic paramagnon couplings, mediated by electronic excitations. Such quartic couplings contain a
product of four fermionic Green's functions and include the effect of collisions between spin fluctuations. When re-summed these diagrams lead to the maximum of the electronic Raman response function at the structural transition in the $B_{1g}$ channel, and the suppression of the response in the $A_{1g}$ channel.

The method that we developed analysed the Raman response function only in the regime of small frequencies. It would be desirable to extend it to the entire frequency range, such that one can analyse the entire shape of the Raman response function as a function of temperature, and possibly be able to extract some information about the dynamical nematic susceptibility. 

Further, one might expect a charge driven nematic phase to have similar signatures in the Raman response function. This could be relevant to the peculiar case of FeSe, where the nematic phase has been detected, but no magnetic phase has been seen. \cite{Anna2, Baek} In order to do so, we would need to develop a theoretical method that goes beyond the large $N$ expansion.

\section{Acknowledgement}
We  acknowledge useful discussions with A. Chubukov and R. Fernandes. U.\,K. acknowledges the support from the Helmholtz Association, through Helmholtz post-doctoral grant PD-075\,  ``Unconventional order and superconductivity in pnictides''. J.\,S., F.\,K., T.\,.B. and R.\,H. acknowledge the support from  Deutsche Forschungsgemeinschaft (DFG) through the Priority Program SPP 1458 ``Hochtemperatur-Supraleitung in Eisenpniktiden'' (project-nos. SCHM 1031/5-1 and HA2071/7-2). Y.\,G. acknowledges financial support from ANR grant PNICTIDES.
\\
Note: In the final stages of the preparations of the manuscript we became aware of arXiv:1504.05054,
where the behaviour of the Raman response function in the vicinity of the structural transition has been analyzed. Where there is overlap with this work, our results agree.
\begin{appendix}
\section{Effective action of the SU(N) fermionic model}

\subsection{Some useful SU(N) identities}
Here, we present some useful identities for the structure constants of $SU(N)$. They have been used to determine the scaling of the boxed Aslamazov-Larkin diagrams with $N$, and to develop the Ginzburg-Landau expansion of the effective action in powers of spin fluctuation fields $\Delta$ (see Sec. \ref{microscopics}). We begin by listing some standard $SU(N)$ identities for the matrices ${\boldsymbol{\lambda}}_{i}$, where $i=1,..N^{2}-1$. All repeated indices are summed over.
\begin{eqnarray}
\{\boldsymbol{\lambda}_{j},\boldsymbol{\lambda}_{k}\} & = & \frac{1}{N}\delta_{jk}+{\mathbf{d}}_{jkl}\boldsymbol{\lambda}_{l}\;\;\; {\mathbf{d}}_{jkl}={\mathbf{d}}_{kjl},
\label{anticom}  \\
\left [\boldsymbol{\lambda}_{j},\boldsymbol{\lambda}_{k} \right ] & = & i{\mathbf{f}}_{jkl}\boldsymbol{\lambda}_{l}\;\;\; {\mathbf{f}}_{jkl}=-{\mathbf{f}}_{kjl},
\label{com}  \\
\boldsymbol{\lambda}_{j}\boldsymbol{\lambda}_{k} & = & \frac{1}{2N}\delta_{jk}+\frac{1}{2}{\mathbf{R}}_{jkl}\boldsymbol{\lambda}^{l},
\label{a} 
\\
{\mathbf{R}}_{jkl} & \coloneqq & {\mathbf{d}}_{jkl}+ i{\mathbf{f}}_{jkl}.
\label{Rdef}
\end{eqnarray}
Here ${\mathbf{d}}_{kjl}$ is symmetric under the exchange of its indices, while ${\mathbf{f}}_{kjl}$ is antisymmetric under the exchange of neighbouring indices.  Further, some useful relations for the summations of structure constants can be derived, \cite{maths1, maths2} which read
\begin{eqnarray}
{\mathbf{d}}_{akl} {\mathbf{d}}_{bkl} &= &\frac{N^{2}-4}{N} \delta_{ab},
\label{dd}  \\
{\mathbf{f}}_{akl} {\mathbf{f}}_{bkl} &= & N \delta_{ab},
\label{ff}  \\
\sum_{i} {\mathbf{d}}_{iij}=0.
\label{d}
\end{eqnarray}

Useful identities that involve the traces of the $SU(N)$ matrices are
\begin{eqnarray}
 \mathrm{Tr}\left (\boldsymbol{\lambda}_{i} \right ) & = & 0,
 \label{b} \\
\mathrm{Tr}\left ( \mathds{1} \right ) & = & N,
\label{c} \\
\mathrm{Tr}\left ( \boldsymbol{\lambda}_{i} \boldsymbol{\lambda}_{j} \right ) & = & \frac{1}{2}\delta_{ij}.
\label{trace2}
\end{eqnarray}
In order to analyse the trace of the product of four $SU(N)$ generators we evaluate
\begin{eqnarray}
\mathrm{Tr}\left (\boldsymbol{\lambda}_{i}\boldsymbol{\lambda}_{j}\boldsymbol{\lambda}_{k}\boldsymbol{\lambda}_{l} \right ) & = & \mathrm{Tr} \left[\left( \frac{1}{2N}\delta_{ij}+\frac{1}{2}{\mathbf{R}}_{ijp}\boldsymbol{\lambda}_{p}\right) \right.
\nonumber \\
& &\times \left. \left( \frac{1}{2N}\delta_{kl}+\frac{1}{2}{\mathbf{R}}_{klr}\boldsymbol{\lambda}_{r}\right )\right]\nonumber \\
 & =& \frac{1}{4N}\delta_{ij}\delta_{kl}+\frac{1}{8}{\mathbf{R}}_{ijp}{\mathbf{R}}_{klp},
 \label{trace4}
\end{eqnarray}
where we used the identity (\ref{a}) in the first line, as well as Eq. (\ref{b}) and Eq. (\ref{trace2}) in the second line.
These results will be of importance for the subsequent analysis of higher order diagrams.
\subsection{Effective action from tr log expansion}
\label{eff_action_derivation}
First we calculate the quadratic terms in the free energy expansion. This is given by
\begin{eqnarray}
\frac{1}{2}{\mathrm{Tr}}\left (\mathcal{G}_{0} \mathcal{V}_{\Delta}\right )^{2} & = &  \sum_{\alpha}
\int_{k} G_{\alpha, k} G_{\Gamma, k} \sum_{i,j=1}^{N^2-1} {\mathrm{Tr}}\left ({\boldsymbol{\lambda}}_{i}{\boldsymbol{\lambda}}_{j} \right)\Delta_{\alpha}^{i}\Delta_{\alpha}^{j}
\nonumber \\
&=& \frac{1}{2}\sum_{\alpha} \int_{k} G_{\alpha, k}G_{\Gamma, k} 
\lvert \boldsymbol{\Delta}_{\alpha} \rvert ^2,
\label{quadratic_action}
\end{eqnarray}
where $\alpha=X,Y$ and we used the identity (\ref{trace2}).

Next we calculate the quartic term in the free energy expansion

\begin{eqnarray}
\frac{1}{4}{\mathrm{Tr}}\left (\mathcal{G}_{0} \mathcal{V}_{\Delta}\right )^{4} &=& \frac{1}{2} 
{\mathrm{Tr}}\left ({\boldsymbol{\lambda}}_{i}{\boldsymbol{\lambda}}_{j}{\boldsymbol{\lambda}}_{k}{\boldsymbol{\lambda}}_{l} \right) \nonumber \\
& &\times \sum_{\alpha=X,Y} g_{\alpha \alpha}
\Delta^{i}_{\alpha} \Delta^{j}_{\alpha} \Delta^{k}_{\alpha} \Delta^{l}_{\alpha}
\nonumber \\
& &+
\frac{1}{2} 
{\mathrm{Tr}}\left ({\boldsymbol{\lambda}}_{i}{\boldsymbol{\lambda}}_{j}{\boldsymbol{\lambda}}_{k}{\boldsymbol{\lambda}}_{l} \right)\nonumber \\
& &\times  \sum_{\alpha=X,Y} g_{\alpha \bar \alpha} \Delta^{i}_{\bar \alpha} \Delta^{j}_{\bar \alpha} \Delta^{k}_{\alpha} \Delta^{l}_{\alpha},
 \nonumber \\
 \label{quartic_0}
\end{eqnarray}
with
\begin{eqnarray}
g_{XX} & = & g_{YY} =\int_{k} G^{2}_{X, k} G^{2}_{\Gamma, k},
\nonumber \\
g_{XY} & = & g_{YX} =\int_{k} G_{X, k} G_{Y, k} G^{2}_{\Gamma, k},
\end{eqnarray}
and we used the notation $\bar \alpha$ for 'not $\alpha$', i.e. if $\alpha=X$ then $\bar \alpha=Y$ and vice versa.
We further substitute the identity (\ref{trace4}) in (\ref{quartic_0}), to write 
\begin{equation}
\frac{1}{4}{\mathrm{Tr}}\left (G \mathcal{V}_{\Delta}\right )^{4} = K_1 + K_2,
\end{equation}
where
\begin{eqnarray}
K_1 &= &\frac{1}{8 N} \sum_{\alpha=X,Y} g_{\alpha \alpha}
\lvert \boldsymbol{\Delta}_{\alpha} \rvert ^4
\nonumber \\
& &+\frac{1}{8 N} \sum_{\alpha=X,Y} g_{\alpha \bar \alpha}
\lvert \boldsymbol{\Delta}_{\alpha} \rvert ^2 \lvert \boldsymbol{\Delta}_{\bar \alpha} \rvert ^2,
\nonumber \\
K_2 &=& \sum_{\alpha=X,Y} \frac{g_{\alpha \alpha}}{16} {\mathbf{R}}_{ijp} {\mathbf{R}}_{klp}  \Delta^{i}_{\alpha} \Delta^{j}_{\alpha} \Delta^{k}_{\alpha} \Delta^{l}_{\alpha}
\nonumber \\
& &+\sum_{\alpha=X,Y} \frac{g_{\alpha \bar \alpha}}{16} {\mathbf{R}}_{ijp} {\mathbf{R}}_{klp} 
\Delta^{i}_{\bar \alpha} \Delta^{j}_{\bar \alpha} \Delta^{k}_{\alpha} \Delta^{l}_{\alpha}.
\label{k1k2}
 \end{eqnarray}
Since $K_2 \sim N^{-5}$, while $K_1 \sim N^{-1}$, the term $K_2$ can be omitted in the large $N$ limit.  

Combining (\ref{k1k2}) and (\ref{quadratic_action}), the effective action in the large $N$ limit can be written as
\begin{equation}
S_{\mathrm{eff}}\left[\boldsymbol{\Delta}_{X},\boldsymbol{\Delta}_{Y}\right]=\sum_{i}r_{0,i}\Delta_{i}^{2}+\sum_{i,j}u_{ij}\Delta_{i}^{2}\Delta_{j}^{2},
\label{effective_S3}
\end{equation}
 with the coefficients:
\begin{eqnarray}
r_{0,i} & = & \frac{2}{u_{\mathrm{s}}}+\frac{1}{2}\int_{k}G_{\Gamma,k}G_{i,k},\nonumber \\
u_{ij} & = & \frac{1}{8 N}\int_{k}G_{\Gamma,k}^{2}G_{i,k}G_{j,k}.
\label{GL}
\end{eqnarray}
We note that in the large $N$ approximation there are no $\boldsymbol{\Delta}_{X} \cdot \boldsymbol{\Delta}_{Y}$ terms in the action; however if one considers corrections to large $N$ these terms might appear in the effective action.

\section{Identities containing products of traces of $SU(N)$ generators}
\label{trace_prefactor}
In this appendix we derive further identities for the traces of the $SU(N)$ generators, which have been used to deduce the dependence of the Aslamazov-Larkin boxed diagrams on $N$.
In particular, we would like to calculate
\begin{eqnarray}
T_{m}& \coloneqq  & {\mathrm{Tr}}\left ({\boldsymbol{\lambda}}_{i_{1}}{\boldsymbol{\lambda}}_{i_{2}} \right ) {\mathrm{Tr}}\left ({\boldsymbol{\lambda}}_{i_{2}}{\boldsymbol{\lambda}}_{i_{1}} 
{\boldsymbol{\lambda}}_{i_{3}}{\boldsymbol{\lambda}}_{i_{4}} \right ) 
\nonumber \\
& &\times 
{\mathrm{Tr}}\left ({\boldsymbol{\lambda}}_{i_{4}}{\boldsymbol{\lambda}}_{i_{3}} 
{\boldsymbol{\lambda}}_{i_{5}}{\boldsymbol{\lambda}}_{i_{6}} \right ) \ldots
\nonumber \\
& &\times {\mathrm{Tr}}\left ({\boldsymbol{\lambda}}_{i_{2m}}{\boldsymbol{\lambda}}_{i_{2m-1}} 
{\boldsymbol{\lambda}}_{i_{2m+1}}{\boldsymbol{\lambda}}_{i_{2m+2}} \right ) \nonumber \\
& &\times {\mathrm{Tr}}\left ({\boldsymbol{\lambda}}_{i_{2m+2}}{\boldsymbol{\lambda}}_{i_{2m+1}} \right ).
\end{eqnarray}
We begin by considering $m=1$. Written out explicitly, it follows:
\begin{eqnarray}
T_{1} & = & {\mathrm{Tr}}({\boldsymbol{\lambda}}_{i}{\boldsymbol{\lambda}}_{j}){\mathrm{Tr}}({\boldsymbol{\lambda}}_{k}{\boldsymbol{\lambda}}_{l}){\mathrm{Tr}}({\boldsymbol{\lambda}}_{j}{\boldsymbol{\lambda}}_{i}{\boldsymbol{\lambda}}_{l}{\boldsymbol{\lambda}}_{k})\nonumber \\
 & = & \left(\frac{1}{4}\delta_{ij}\delta_{kl}\right)\left(\frac{1}{4N}\delta_{ij}\delta_{kl}+\frac{1}{8}{\mathbf{R}}_{jir}{\mathbf{R}}_{lkr}\right)\nonumber \\
 & = & \frac{1}{4}\frac{1}{(4N)}\sum_{ijkl}\delta_{ij}\delta_{kl}+\sum_{ikr}\frac{1}{32}{\mathbf{R}}_{iir}{\mathbf{R}}_{kkr}\nonumber \\
 & =& \frac{1}{4}\frac{1}{(4N)}\left (N^{2}-1 \right )^{2},
 \label{t1}
\end{eqnarray}
where we have used (\ref{trace2}) and (\ref{trace4}) to get to the second line, and the fact that ${\mathbf{R}}_{iir}=0$ in the penultimate line, which is a consequence of (\ref{d}) and the antisymmetry of ${\mathbf{f}}$.
Using the same set of identities, we find that 
\begin{eqnarray}
T_{2} &= & {\mathrm{Tr}}({\boldsymbol{\lambda}}_{i}{\boldsymbol{\lambda}}_{j}){\mathrm{Tr}}({\boldsymbol{\lambda}}_{k}{\boldsymbol{\lambda}}_{l}){\mathrm{Tr}}({\boldsymbol{\lambda}}_{j}{\boldsymbol{\lambda}}_{i}{\boldsymbol{\lambda}}_{s}{\boldsymbol{\lambda}}_{r})
 {\mathrm{Tr}}({\boldsymbol{\lambda}}_{r}{\boldsymbol{\lambda}}_{s}{\boldsymbol{\lambda}}_{l}{\boldsymbol{\lambda}}_{k})\nonumber \\
 &=& \left(\frac{1}{4}\delta_{ij}\delta_{kl}\right)\left(\frac{1}{4N}\delta_{ij}\delta_{sr}+\frac{1}{8}{\mathbf{R}}_{jit}{\mathbf{R}}_{srt}\right)
 \nonumber \\
& \times & \left(\frac{1}{4N}\delta_{sr}\delta_{kl}+\frac{1}{8}{\mathbf{R}}_{rsz}{\mathbf{R}}_{lkz}\right)\nonumber \\
 &= & \frac{1}{4}\left(\frac{1}{4N}\right)^{2}\sum_{ijklsr}\delta_{ij}\delta_{kl}\delta_{sr}\nonumber \\
&= & \frac{1}{4}\left(\frac{1}{4N}\right)^{2}(N^{2}-1)^{3} .
\label{t2}
\end{eqnarray}
Similarly, one can deduce that 
\begin{equation}
T_{m}= \frac{1}{4}\left(\frac{1}{4N}\right)^{m}(N^{2}-1)^{m+1}
\approx \frac{N^2}{4} \left ( \frac{N}{4} \right )^m.
\label{tm}
\end{equation}
\end{appendix}


\begin{thebibliography}{0}
\expandafter\ifx\csname natexlab\endcsname\relax\def\natexlab#1{#1}\fi
\expandafter\ifx\csname bibnamefont\endcsname\relax
  \def\bibnamefont#1{#1}\fi
\expandafter\ifx\csname bibfnamefont\endcsname\relax
  \def\bibfnamefont#1{#1}\fi
\expandafter\ifx\csname citenamefont\endcsname\relax
  \def\citenamefont#1{#1}\fi
\expandafter\ifx\csname url\endcsname\relax
  \def\url#1{\texttt{#1}}\fi
\expandafter\ifx\csname urlprefix\endcsname\relax\def\urlprefix{URL }\fi
\providecommand{\bibinfo}[2]{#2}
\providecommand{\eprint}[2][]{\url{#2}}




\bibitem[{\citenamefont{Paglione et~al.}(2010)\citenamefont{Paglione, and Greene}}]{review}
\bibinfo{author}{\bibfnamefont{J.} \bibnamefont{Paglione}},  \bibnamefont{and}
  \bibinfo{author}{\bibfnamefont{R.~L.} \bibnamefont{Greene}},
   \bibinfo{journal}{Nature Phys.}
  \textbf{\bibinfo{volume}{6}}, \bibinfo{pages}{645} (\bibinfo{year}{2010}).

\bibitem[{\citenamefont{Kim et~al.}(2011)\citenamefont{Kim, Fernandes, Kreyssig, Kim,
Thaler, Bud'ko, Canfield, McQueeney, Schmalian,
and Goldman}}]{Kim11}
  \bibinfo{author}{\bibfnamefont{M.~G.} \bibnamefont{Kim}},
  \bibinfo{author}{\bibfnamefont{R.~M.} \bibnamefont{Fernandes}},
  \bibinfo{author}{\bibfnamefont{A.} \bibnamefont{Kreyssig}},
 \bibinfo{author}{\bibfnamefont{J.~W.} \bibnamefont{Kim}}, 
 \bibinfo{author}{\bibfnamefont{A.} \bibnamefont{Thaler}},  
 \bibinfo{author}{\bibfnamefont{S.~L.} \bibnamefont{Bud'ko}},  
\bibinfo{author}{\bibfnamefont{P.~C.} \bibnamefont{Canfield}}, 
\bibinfo{author}{\bibfnamefont{R.~J.} \bibnamefont{McQueeney}},  
\bibinfo{author}{\bibfnamefont{J.} \bibnamefont{Schmalian}},    \bibnamefont{and}
 \bibinfo{author}{\bibfnamefont{A.~I.} \bibnamefont{Goldman}}, 
   \bibinfo{journal}{Phys. Rev. B}
  \textbf{\bibinfo{volume}{83}}, \bibinfo{pages}{134522} (\bibinfo{year}{2011}).

\bibitem[{\citenamefont{Fernandes et~al.}(2011)\citenamefont{Fernandes, and Schmalian}}]{nematic_review}
  \bibinfo{author}{\bibfnamefont{R.~M.} \bibnamefont{Fernandes}},  \bibnamefont{and}
  \bibinfo{author}{\bibfnamefont{J.} \bibnamefont{Schmalian}},
   \bibinfo{journal}{Supercond. Sci. Technol.}
  \textbf{\bibinfo{volume}{25}}, \bibinfo{pages}{ 084005} (\bibinfo{year}{2012}).

\bibitem[{\citenamefont{Birgenau~al.}(2011)\citenamefont{Fernandes, and Schmalian}}]{Birgeneau11}
  \bibinfo{author}{\bibfnamefont{C.~R.} \bibnamefont{Rotundu}},  \bibnamefont{and}
  \bibinfo{author}{\bibfnamefont{R.~J.} \bibnamefont{Birgeneau}},
   \bibinfo{journal}{Phys. Rev. B}
  \textbf{\bibinfo{volume}{84}}, \bibinfo{pages}{092501} (\bibinfo{year}{2011}).

\bibitem[{\citenamefont{Kim et~al.}(2012)\citenamefont{Kasahara, Shi, Hashimoto, Tonegawa,
 Mizukami, Shibauchi, Sugimoto, Fukuda, Terashima, Nevidomskyy, and Matsuda,}}]{matsuda_t}
  \bibinfo{author}{\bibfnamefont{S.} \bibnamefont{Kasahara}},
  \bibinfo{author}{\bibfnamefont{H.~J.} \bibnamefont{Shi}},
  \bibinfo{author}{\bibfnamefont{K.} \bibnamefont{Hashimoto}},
 \bibinfo{author}{\bibfnamefont{S.} \bibnamefont{Tonegawa}}, 
 \bibinfo{author}{\bibfnamefont{Y.} \bibnamefont{Mizukami}},  
 \bibinfo{author}{\bibfnamefont{T.} \bibnamefont{Shibauchi}},  
\bibinfo{author}{\bibfnamefont{K.} \bibnamefont{Sugimoto}}, 
\bibinfo{author}{\bibfnamefont{T.} \bibnamefont{Fukuda}},  
\bibinfo{author}{\bibfnamefont{T.} \bibnamefont{Terashima}},   
 \bibinfo{author}{\bibfnamefont{A.~H.} \bibnamefont{Nevidomskyy}},  \bibnamefont{and}
  \bibinfo{author}{\bibfnamefont{Y.} \bibnamefont{Matsuda}}, 
   \bibinfo{journal}{Nature}
  \textbf{\bibinfo{volume}{486}}, \bibinfo{pages}{382} (\bibinfo{year}{2012}).

\bibitem[{\citenamefont{Fernandes et~al.}(2014)\citenamefont{Fernandes, Chubukov, and Schmalian}}]{naturereview}
  \bibinfo{author}{\bibfnamefont{R.~M.} \bibnamefont{Fernandes}},
    \bibinfo{author}{\bibfnamefont{A.~V.} \bibnamefont{Chubukov}},  \bibnamefont{and}
  \bibinfo{author}{\bibfnamefont{J.} \bibnamefont{Schmalian}},
   \bibinfo{journal}{Nature Phys.}
  \textbf{\bibinfo{volume}{10}}, \bibinfo{pages}{97-104} (\bibinfo{year}{2014}).


\bibitem[{\citenamefont{Fernandes et~al.}(2012)\citenamefont{Fernandes, Chubukov, Knolle, Eremin, and Schmalian}}]{Fernandes12}
  \bibinfo{author}{\bibfnamefont{R.~M.} \bibnamefont{Fernandes}},
    \bibinfo{author}{\bibfnamefont{A.~V.} \bibnamefont{Chubukov}},
  \bibinfo{author}{\bibfnamefont{J.} \bibnamefont{Knolle}},
  \bibinfo{author}{\bibfnamefont{I.} \bibnamefont{Eremin}},  \bibnamefont{and}
  \bibinfo{author}{\bibfnamefont{J.} \bibnamefont{Schmalian}},
   \bibinfo{journal}{Phys. Rev. B}
  \textbf{\bibinfo{volume}{85}}, \bibinfo{pages}{024534} (\bibinfo{year}{2012}).

\bibitem[{\citenamefont{Xu et~al.}(2008)\citenamefont{Xu, M\"uller, and S. Sachdev}}]{Xu08}
  \bibinfo{author}{\bibfnamefont{C.} \bibnamefont{Xu}},
    \bibinfo{author}{\bibfnamefont{M.} \bibnamefont{M\"uller}},  \bibnamefont{and}
  \bibinfo{author}{\bibfnamefont{S.} \bibnamefont{Sachdev}},
   \bibinfo{journal}{Phys. Rev. B}
  \textbf{\bibinfo{volume}{78}}, \bibinfo{pages}{020501(R)} (\bibinfo{year}{2008}).

\bibitem[{\citenamefont{Fang et~al.}(2008)\citenamefont{Fang, Yao, Tsai, Hu, and Kivelson}}]{Fang08}
  \bibinfo{author}{\bibfnamefont{C.} \bibnamefont{Fang}},
    \bibinfo{author}{\bibfnamefont{H.} \bibnamefont{Yao}},
  \bibinfo{author}{\bibfnamefont{W.~F.} \bibnamefont{Tsai}},
    \bibinfo{author}{\bibfnamefont{J.~P.} \bibnamefont{Hu}},  \bibnamefont{and}
\bibinfo{author}{\bibfnamefont{S.~A.} \bibnamefont{Kivelson}},
   \bibinfo{journal}{Phys. Rev. B}
  \textbf{\bibinfo{volume}{77}}, \bibinfo{pages}{224509} (\bibinfo{year}{2008}).

\bibitem[{\citenamefont{Qi et~al.}(2009)\citenamefont{Qi, and Xu}}]{Qi09}
  \bibinfo{author}{\bibfnamefont{Y.} \bibnamefont{Qi}},  \bibnamefont{and}
    \bibinfo{author}{\bibfnamefont{C.} \bibnamefont{Xu}},
   \bibinfo{journal}{Phys. Rev. B}
  \textbf{\bibinfo{volume}{80}}, \bibinfo{pages}{094402} (\bibinfo{year}{2009}).

\bibitem[{\citenamefont{Cano et~al.}(2010)\citenamefont{Cano, Civelli, Eremin, and Paul}}]{Cano10}
  \bibinfo{author}{\bibfnamefont{A.} \bibnamefont{Cano}},
    \bibinfo{author}{\bibfnamefont{M.} \bibnamefont{Civelli}},
  \bibinfo{author}{\bibfnamefont{I.} \bibnamefont{Eremin}},  \bibnamefont{and}
    \bibinfo{author}{\bibfnamefont{I.} \bibnamefont{Paul}},
   \bibinfo{journal}{Phys. Rev. B}
  \textbf{\bibinfo{volume}{82}}, \bibinfo{pages}{020408(R)} (\bibinfo{year}{2010}).

\bibitem[{\citenamefont{Fernandes et~al.}(2010)\citenamefont{Fernandes, VanBebber, Bhattacharya, Chandra, Keppens, Mandrus, McGuire, Sales, Sefat, and Schmalian}}]{Fernandes2010}
  \bibinfo{author}{\bibfnamefont{R.~M.} \bibnamefont{Fernandes}},
    \bibinfo{author}{\bibfnamefont{L.~H.} \bibnamefont{VanBebber}},
  \bibinfo{author}{\bibfnamefont{S.} \bibnamefont{Bhattacharya}},
    \bibinfo{author}{\bibfnamefont{P.} \bibnamefont{Chandra}},
    \bibinfo{author}{\bibfnamefont{V.} \bibnamefont{Keppens}},
    \bibinfo{author}{\bibfnamefont{D.} \bibnamefont{Mandrus}},
    \bibinfo{author}{\bibfnamefont{M.~A.} \bibnamefont{McGuire}},
        \bibinfo{author}{\bibfnamefont{B.~C.} \bibnamefont{Sales}},
        \bibinfo{author}{\bibfnamefont{A.~S.} \bibnamefont{Sefat}},  \bibnamefont{and}
        \bibinfo{author}{\bibfnamefont{J.} \bibnamefont{Schmalian}},
   \bibinfo{journal}{Phys.
Rev. Lett.}
  \textbf{\bibinfo{volume}{105}}, \bibinfo{pages}{157003} (\bibinfo{year}{2010}).

\bibitem[{\citenamefont{Tanatar et~al.}(2010)\citenamefont{Tanatar, Blomberg, Kreyssig, Kim, Ni, Thaler, Bud'ko, Canfield, Goldman, Mazin, and Prozorov}}]{Tanatar10}
  \bibinfo{author}{\bibfnamefont{M.~A.} \bibnamefont{Tanatar}},
    \bibinfo{author}{\bibfnamefont{E.~C.} \bibnamefont{Blomberg}},
  \bibinfo{author}{\bibfnamefont{A.} \bibnamefont{Kreyssig}},
    \bibinfo{author}{\bibfnamefont{M.~G.} \bibnamefont{Kim}},
    \bibinfo{author}{\bibfnamefont{N.} \bibnamefont{Ni}},
    \bibinfo{author}{\bibfnamefont{A.} \bibnamefont{Thaler}},
    \bibinfo{author}{\bibfnamefont{S.~L.} \bibnamefont{ Bud'ko}},
        \bibinfo{author}{\bibfnamefont{P.~C.} \bibnamefont{Canfield}},
        \bibinfo{author}{\bibfnamefont{A.~I.} \bibnamefont{Goldman}},
        \bibinfo{author}{\bibfnamefont{I.~I.} \bibnamefont{Mazin}},  \bibnamefont{and}
          \bibinfo{author}{\bibfnamefont{R.} \bibnamefont{Prozorov}},
   \bibinfo{journal}{Phys.
Rev. B}
  \textbf{\bibinfo{volume}{81}}, \bibinfo{pages}{184508} (\bibinfo{year}{2010}).

\bibitem[{\citenamefont{Chu et~al.}(2012)\citenamefont{Chu, Kuo, Analytis, and Fisher}}]{Chu2012}
  \bibinfo{author}{\bibfnamefont{J.~H.} \bibnamefont{Chu}},
   \bibinfo{author}{\bibfnamefont{H.~H.} \bibnamefont{Kuo}},
  \bibinfo{author}{\bibfnamefont{J.~G.} \bibnamefont{Analytis}},  \bibnamefont{and}
    \bibinfo{author}{\bibfnamefont{I.~R.} \bibnamefont{Fisher}},
   \bibinfo{journal}{Science}
  \textbf{\bibinfo{volume}{337}}, \bibinfo{pages}{719} (\bibinfo{year}{2012}).

\bibitem[{\citenamefont{Chu et~al.}(2010)\citenamefont{Chu, Analytis, De Greve, McMahon, Islam, Yamamoto, and Fisher}}]{Chu10}
  \bibinfo{author}{\bibfnamefont{J.~H.} \bibnamefont{Chu}},
  \bibinfo{author}{\bibfnamefont{J.~G.} \bibnamefont{Analytis}},
  \bibinfo{author}{\bibfnamefont{K.} \bibnamefont{De Greve}},
  \bibinfo{author}{\bibfnamefont{P.~L.} \bibnamefont{McMahon}},
  \bibinfo{author}{\bibfnamefont{Z.} \bibnamefont{Islam}},
    \bibinfo{author}{\bibfnamefont{Y.} \bibnamefont{Yamamoto}},  \bibnamefont{and}
    \bibinfo{author}{\bibfnamefont{I.~R.} \bibnamefont{Fisher}},
   \bibinfo{journal}{Science}
  \textbf{\bibinfo{volume}{329}}, \bibinfo{pages}{824} (\bibinfo{year}{2010}).


\bibitem[{\citenamefont{Jiang et~al.}(2013)\citenamefont{Jiang, Jeevan, Dong, and P. Gegenwart}}]{Jiang2013}
  \bibinfo{author}{\bibfnamefont{Shuai} \bibnamefont{Jiang}},
  \bibinfo{author}{\bibfnamefont{H.~S.} \bibnamefont{Jeevan}},
  \bibinfo{author}{\bibfnamefont{Jinkui} \bibnamefont{Dong}},  \bibnamefont{and}
  \bibinfo{author}{\bibfnamefont{P.} \bibnamefont{Gegenwart}},
   \bibinfo{journal}{Phys. Rev. Lett.}
  \textbf{\bibinfo{volume}{110}}, \bibinfo{pages}{067001} (\bibinfo{year}{2013}).

\bibitem[{\citenamefont{Jiang et~al.}(2011)\citenamefont{Dusza, Lucarelli, Pfuner, Chu,
Fisher, and Degiorgi}}]{Dusza2011}
  \bibinfo{author}{\bibfnamefont{A.} \bibnamefont{Dusza}},
  \bibinfo{author}{\bibfnamefont{A.} \bibnamefont{Lucarelli}},
  \bibinfo{author}{\bibfnamefont{F.} \bibnamefont{Pfuner}},
  \bibinfo{author}{\bibfnamefont{J.~H.} \bibnamefont{Chu}},
  \bibinfo{author}{\bibfnamefont{I.~R.} \bibnamefont{Fisher}},  \bibnamefont{and}
  \bibinfo{author}{\bibfnamefont{L.} \bibnamefont{Degiorgi}},
   \bibinfo{journal}{Europhys. Lett.}
  \textbf{\bibinfo{volume}{93}}, \bibinfo{pages}{37002} (\bibinfo{year}{2011}).

\bibitem[{\citenamefont{Nakajima et~al.}(2011)\citenamefont{Nakajima, Liang, Ishida, Tomioka, Kihou, Lee, Iyo, Eisaki, Kakeshita, Ito, and Uchida}}]{Nakajima2011}
  \bibinfo{author}{\bibfnamefont{M.} \bibnamefont{Nakajima}},
  \bibinfo{author}{\bibfnamefont{T.} \bibnamefont{Liang}},
  \bibinfo{author}{\bibfnamefont{S.} \bibnamefont{Ishida}},
  \bibinfo{author}{\bibfnamefont{Y.} \bibnamefont{Tomioka}},
    \bibinfo{author}{\bibfnamefont{K.} \bibnamefont{Kihou}},
  \bibinfo{author}{\bibfnamefont{C.~H.} \bibnamefont{Lee}},
  \bibinfo{author}{\bibfnamefont{A.} \bibnamefont{Iyo}},
    \bibinfo{author}{\bibfnamefont{H.} \bibnamefont{Eisaki}},
    \bibinfo{author}{\bibfnamefont{T.} \bibnamefont{Kakeshita}},
    \bibinfo{author}{\bibfnamefont{T.} \bibnamefont{Ito}},  \bibnamefont{and}
    \bibinfo{author}{\bibfnamefont{S.} \bibnamefont{Uchida}},
   \bibinfo{journal}{Proc. Natl. Acad. Sci. U.S.A.}
  \textbf{\bibinfo{volume}{108}}, \bibinfo{pages}{12 238} (\bibinfo{year}{2011}).

\bibitem[{\citenamefont{Rosenthal et~al.}(2014)\citenamefont{Rosenthal, Andrade, Arguello,
Fernandes, Xing, Wang, Jin, Millis, and Pasupathy}}]{Rosenthal13}
  \bibinfo{author}{\bibfnamefont{E.~P.} \bibnamefont{Rosenthal}},
  \bibinfo{author}{\bibfnamefont{E.~F.} \bibnamefont{Andrade}},
  \bibinfo{author}{\bibfnamefont{C.~J.} \bibnamefont{Arguello}},
  \bibinfo{author}{\bibfnamefont{R.~M.} \bibnamefont{Fernandes}},
    \bibinfo{author}{\bibfnamefont{L.~Y.} \bibnamefont{Xing}},
  \bibinfo{author}{\bibfnamefont{X.~C.} \bibnamefont{Wang}},
  \bibinfo{author}{\bibfnamefont{C.~Q.} \bibnamefont{Jin}},
    \bibinfo{author}{\bibfnamefont{A.~J.} \bibnamefont{Millis}},  \bibnamefont{and}
    \bibinfo{author}{\bibfnamefont{A.~N.} \bibnamefont{Pasupathy}},
   \bibinfo{journal}{Nature Phys.}
  \textbf{\bibinfo{volume}{10}}, \bibinfo{pages}{225–232} (\bibinfo{year}{2014}).

\bibitem[{\citenamefont{Fernandes et~al.}(2013)\citenamefont{Fernandes, B\"ohmer, Meingast,
and Schmalian}}]{Fernandes13_shear}
  \bibinfo{author}{\bibfnamefont{R.~M.} \bibnamefont{Fernandes}},
  \bibinfo{author}{\bibfnamefont{A.~E.} \bibnamefont{B\"ohmer}},
  \bibinfo{author}{\bibfnamefont{C.} \bibnamefont{Meingast}},  \bibnamefont{and}
  \bibinfo{author}{\bibfnamefont{J.} \bibnamefont{Schmalian}},
   \bibinfo{journal}{Phys. Rev. Lett.}
  \textbf{\bibinfo{volume}{111}}, \bibinfo{pages}{137001} (\bibinfo{year}{2013}).

\bibitem[{\citenamefont{Kontani et~al.}(2014)\citenamefont{Kontani, and Yamakawa}}]{Kontani1}
  \bibinfo{author}{\bibfnamefont{H.} \bibnamefont{Kontani}},  \bibnamefont{and}
  \bibinfo{author}{\bibfnamefont{Y.} \bibnamefont{Yamakawa}},
   \bibinfo{journal}{Phys. Rev. Lett.}
  \textbf{\bibinfo{volume}{113}}, \bibinfo{pages}{047001} (\bibinfo{year}{2014}).

\bibitem[{\citenamefont{Kontani et~al.}(2011)\citenamefont{Kontani, Saito, and Onari}}]{Kontani2}
  \bibinfo{author}{\bibfnamefont{H.} \bibnamefont{Kontani}},
  \bibinfo{author}{\bibfnamefont{T.} \bibnamefont{Saito}},  \bibnamefont{and}
    \bibinfo{author}{\bibfnamefont{S.} \bibnamefont{Onari}},
   \bibinfo{journal}{Phys. Rev. B}
  \textbf{\bibinfo{volume}{84}}, \bibinfo{pages}{024528} (\bibinfo{year}{2011}).

\bibitem[{\citenamefont{B\"ohmer et~al.}(2014)\citenamefont{B\"ohmer, Burger, Hardy, Wolf, Schweiss, Fromknecht, Reinecker, Schranz, and Meingast}}]{Anna1}
  \bibinfo{author}{\bibfnamefont{A.~E.} \bibnamefont{B\"ohmer}},
  \bibinfo{author}{\bibfnamefont{P.} \bibnamefont{Burger}},
  \bibinfo{author}{\bibfnamefont{F.} \bibnamefont{Hardy}},
  \bibinfo{author}{\bibfnamefont{T.} \bibnamefont{Wolf}},
  \bibinfo{author}{\bibfnamefont{P.} \bibnamefont{Schweiss}},
  \bibinfo{author}{\bibfnamefont{R.} \bibnamefont{Fromknecht}},
  \bibinfo{author}{\bibfnamefont{M.} \bibnamefont{Reinecker}},
  \bibinfo{author}{\bibfnamefont{W.} \bibnamefont{Schranz}},  \bibnamefont{and}
  \bibinfo{author}{\bibfnamefont{C.} \bibnamefont{Meingast}},
   \bibinfo{journal}{Phys. Rev. Lett.}
  \textbf{\bibinfo{volume}{112}}, \bibinfo{pages}{047001} (\bibinfo{year}{2014}).

\bibitem[{\citenamefont{B\"ohmer et~al.}(2015)\citenamefont{B\"ohmer, Arai, Hardy, Hattori, Iye, Wolf,  v. L\"ohneysen, Ishida, and Meingast}}]{Anna2}
  \bibinfo{author}{\bibfnamefont{A.~E.} \bibnamefont{B\"ohmer}},
  \bibinfo{author}{\bibfnamefont{T.} \bibnamefont{Arai}},
  \bibinfo{author}{\bibfnamefont{F.} \bibnamefont{Hardy}},
  \bibinfo{author}{\bibfnamefont{T.} \bibnamefont{Hattori}},
  \bibinfo{author}{\bibfnamefont{T.} \bibnamefont{Iye}},
  \bibinfo{author}{\bibfnamefont{T.} \bibnamefont{Wolf}},
  \bibinfo{author}{\bibfnamefont{H.} \bibnamefont{v. L\"ohneysen}},
  \bibinfo{author}{\bibfnamefont{K.} \bibnamefont{Ishida}},  \bibnamefont{and}
  \bibinfo{author}{\bibfnamefont{C.} \bibnamefont{Meingast}},
   \bibinfo{journal}{Phys. Rev. Lett.}
  \textbf{\bibinfo{volume}{114}}, \bibinfo{pages}{027001} (\bibinfo{year}{2015}).

\bibitem[{\citenamefont{Gallais et~al.}(2013)\citenamefont{Gallais, Fernandes, Paul, Chauviere, Yang, Méasson, Cazayous, Sacuto, Colson, and Forget}}]{Gallais}
  \bibinfo{author}{\bibfnamefont{Y.} \bibnamefont{Gallais}},
    \bibinfo{author}{\bibfnamefont{R.~M.} \bibnamefont{Fernandes}},
      \bibinfo{author}{\bibfnamefont{I.} \bibnamefont{Paul}},
  \bibinfo{author}{\bibfnamefont{L.} \bibnamefont{Chauviere}},
      \bibinfo{author}{\bibfnamefont{Y.~X.} \bibnamefont{Yang}},
  \bibinfo{author}{\bibfnamefont{M.~A.} \bibnamefont{Measson}},
  \bibinfo{author}{\bibfnamefont{M.} \bibnamefont{Cazayous}},
    \bibinfo{author}{\bibfnamefont{A.} \bibnamefont{Sacuto}},
        \bibinfo{author}{\bibfnamefont{D.} \bibnamefont{Colson}},  \bibnamefont{and}
        \bibinfo{author}{\bibfnamefont{A.} \bibnamefont{Forget}},
   \bibinfo{journal}{Phys. Rev. Lett.}
 \textbf{\bibinfo{volume}{111}}, \bibinfo{pages}{267001} 
  (\bibinfo{year}{2013}).

\bibitem[{\citenamefont{Zhang et~al.}(2015)\citenamefont{Zhang, Richard, Ding, Sefat, Gillett, Sebastian, Khodas, G. Blumberg}}]{Blumberg1}
  \bibinfo{author}{\bibfnamefont{W.~L.} \bibnamefont{Zhang}},
  \bibinfo{author}{\bibfnamefont{P.} \bibnamefont{Richard}},
  \bibinfo{author}{\bibfnamefont{H.} \bibnamefont{Ding}},
  \bibinfo{author}{\bibfnamefont{Athena S.} \bibnamefont{Sefat}},
    \bibinfo{author}{\bibfnamefont{J.} \bibnamefont{Gillett}},
  \bibinfo{author}{\bibfnamefont{Suchitra E.} \bibnamefont{Sebastian}},
  \bibinfo{author}{\bibfnamefont{M.} \bibnamefont{Khodas}},  \bibnamefont{and}
  \bibinfo{author}{\bibfnamefont{G.} \bibnamefont{Blumberg}},
   \bibinfo{journal}{arXiv:1410.6452}
  (\bibinfo{year}{2015}).

\bibitem[{\citenamefont{Thorsmolle et~al.}(2015)\citenamefont{Thorsmolle, Khodas, Yin, Zhang, Carr, Dai, and G. Blumberg}}]{Blumberg2}
  \bibinfo{author}{\bibfnamefont{V.~K.} \bibnamefont{Thorsmolle}},
    \bibinfo{author}{\bibfnamefont{M.} \bibnamefont{Khodas}},
      \bibinfo{author}{\bibfnamefont{Z.~P.} \bibnamefont{Yin}},
  \bibinfo{author}{\bibfnamefont{Chenglin} \bibnamefont{Zhang}},
      \bibinfo{author}{\bibfnamefont{S.~V.} \bibnamefont{Carr}},
  \bibinfo{author}{\bibfnamefont{Pengcheng} \bibnamefont{Dai}},  \bibnamefont{and}
  \bibinfo{author}{\bibfnamefont{G.} \bibnamefont{Blumberg}},
   \bibinfo{journal}{arXiv:1410.6452}
  (\bibinfo{year}{2015}).

\bibitem[{\citenamefont{Caprara et~al.}(2005)\citenamefont{Caprara, Di Castro, Grilli, and Suppa}}]{Caprara05}
  \bibinfo{author}{\bibfnamefont{S.} \bibnamefont{Caprara}},
    \bibinfo{author}{\bibfnamefont{C.} \bibnamefont{Di Castro}}, 
        \bibinfo{author}{\bibfnamefont{M.} \bibnamefont{Grilli}},  \bibnamefont{and}
    \bibinfo{author}{\bibfnamefont{D.} \bibnamefont{Suppa}},  
   \bibinfo{journal}{Phys. Rev. Lett.}
\textbf{\bibinfo{volume}{95}}, \bibinfo{pages}{117004} 
  (\bibinfo{year}{2005}).
  
  \bibitem[{\citenamefont{Kretzschmar et~al.}(2015)\citenamefont{Kretzschmar, Boehm, Schmalian, Karahasanovic, and Hackl}}]{Rudi}
  \bibinfo{author}{\bibfnamefont{F.} \bibnamefont{Kretzschmar}},
    \bibinfo{author}{\bibfnamefont{T.} \bibnamefont{B\"ohm}},
  \bibinfo{author}{\bibfnamefont{U.} \bibnamefont{Karahasanovic}},
      \bibinfo{author}{\bibfnamefont{J.} \bibnamefont{Schmalian}},  \bibnamefont{and}
       \bibinfo{author}{\bibfnamefont{R.} \bibnamefont{Hackl}},     
   \bibinfo{journal}{(unpublished)}
  (\bibinfo{year}{2015}).

\bibitem[{\citenamefont{Gallais et~al.}(2015)\citenamefont{Gallais, Paul, Chauviere, and Schmalian}}]{Indranil}
  \bibinfo{author}{\bibfnamefont{Y.} \bibnamefont{Gallais}},
    \bibinfo{author}{\bibfnamefont{I.} \bibnamefont{Paul}},
  \bibinfo{author}{\bibfnamefont{L.} \bibnamefont{Chauvière}},  \bibnamefont{and}
      \bibinfo{author}{\bibfnamefont{J.} \bibnamefont{Schmalian}},
   \bibinfo{journal}{arXiv:1504.04570}
  (\bibinfo{year}{2015}).

\bibitem[{\citenamefont{Applegate et~al.}(2011)\citenamefont{Thorsmolle, Khodas, Yin, Zhang, Carr, Dai, and G. Blumberg}}]{Applegate11}
  \bibinfo{author}{\bibfnamefont{R.} \bibnamefont{Applegate}},
    \bibinfo{author}{\bibfnamefont{R.~R.~P.} \bibnamefont{Singh}},
      \bibinfo{author}{\bibfnamefont{C.~C.} \bibnamefont{Chen}},  \bibnamefont{and}
  \bibinfo{author}{\bibfnamefont{T.~P.} \bibnamefont{Devereaux}},
   \bibinfo{journal}{Phys. Rev. B}
 \textbf{\bibinfo{volume}{85}}, \bibinfo{pages}{054411} 
  (\bibinfo{year}{2012}).

\bibitem[{\citenamefont{Lv et~al.}(2011)\citenamefont{Lv, and Phillips}}]{Phillips11}
  \bibinfo{author}{\bibfnamefont{W.} \bibnamefont{Lv}},  \bibnamefont{and}
    \bibinfo{author}{\bibfnamefont{P.} \bibnamefont{Phillips}},
   \bibinfo{journal}{Phys. Rev. B}
 \textbf{\bibinfo{volume}{84}}, \bibinfo{pages}{174512} 
  (\bibinfo{year}{2011}).

\bibitem[{\citenamefont{Liang et~al.}(2013)\citenamefont{Liang, Moreo, and Dagotto}}]{Dagotto13}
  \bibinfo{author}{\bibfnamefont{S.} \bibnamefont{Liang}},
    \bibinfo{author}{\bibfnamefont{A.} \bibnamefont{Moreo}},  \bibnamefont{and}
    \bibinfo{author}{\bibfnamefont{E.} \bibnamefont{Dagotto}},    
   \bibinfo{journal}{Phys. Rev. Lett.}
 \textbf{\bibinfo{volume}{111}}, \bibinfo{pages}{047004} 
  (\bibinfo{year}{2013}).


\bibitem[{\citenamefont{Lee et~al.}(2009)\citenamefont{Lee, Yin, and Ku}}]{w_ku10}
  \bibinfo{author}{\bibfnamefont{C.~C} \bibnamefont{Lee}},
    \bibinfo{author}{\bibfnamefont{W.~G.} \bibnamefont{Yin}},  \bibnamefont{and}
    \bibinfo{author}{\bibfnamefont{W.} \bibnamefont{Ku}},    
   \bibinfo{journal}{Phys. Rev. Lett.}
 \textbf{\bibinfo{volume}{103}}, \bibinfo{pages}{267001} 
  (\bibinfo{year}{2009}).

\bibitem[{\citenamefont{ Kr\"uger et~al.}(2009)\citenamefont{Kr\"uger, Kumar, Zaanen, and van den Brink}}]{kruger1}
  \bibinfo{author}{\bibfnamefont{F.} \bibnamefont{Kr\"uger}},
    \bibinfo{author}{\bibfnamefont{S.} \bibnamefont{Kumar}},
    \bibinfo{author}{\bibfnamefont{J.} \bibnamefont{Zaanen}},    \bibnamefont{and}
        \bibinfo{author}{\bibfnamefont{J.} \bibnamefont{van den Brink}},     
   \bibinfo{journal}{Phys. Rev. B}
 \textbf{\bibinfo{volume}{79}}, \bibinfo{pages}{054504} 
  (\bibinfo{year}{2009}).

\bibitem[{\citenamefont{Lv et~al.}(2010)\citenamefont{Lv, Kr\"uger, and Phillips}}]{kruger2}
    \bibinfo{author}{\bibfnamefont{W.} \bibnamefont{Lv}},
  \bibinfo{author}{\bibfnamefont{F.} \bibnamefont{Kr\"uger}},  \bibnamefont{and}
    \bibinfo{author}{\bibfnamefont{P.} \bibnamefont{Phillips}},   
   \bibinfo{journal}{Phys. Rev. B}
 \textbf{\bibinfo{volume}{82}}, \bibinfo{pages}{045125} 
  (\bibinfo{year}{2010}).

\bibitem[{\citenamefont{Dai et~al.}(2012)\citenamefont{Dai, Hu, and Dagotto}}]{Dai_review}
    \bibinfo{author}{\bibfnamefont{P.} \bibnamefont{Dai}},
  \bibinfo{author}{\bibfnamefont{J.} \bibnamefont{Hu}},  \bibnamefont{and}
    \bibinfo{author}{\bibfnamefont{E.} \bibnamefont{Dagotto}},   
   \bibinfo{journal}{Nature Phys.}
 \textbf{\bibinfo{volume}{8}}, \bibinfo{pages}{709-718} 
  (\bibinfo{year}{2012}).

\bibitem[{\citenamefont{Fang et~al.}(2008)\citenamefont{Fang, Yao, Tsai, Hu, and
Kivelson}}]{Kivelson}
    \bibinfo{author}{\bibfnamefont{C.} \bibnamefont{Fang}},
  \bibinfo{author}{\bibfnamefont{H.} \bibnamefont{Yao}},
    \bibinfo{author}{\bibfnamefont{W.~F.} \bibnamefont{Tsai}},   
        \bibinfo{author}{\bibfnamefont{J.~P.} \bibnamefont{Hu}},    \bibnamefont{and}
    \bibinfo{author}{\bibfnamefont{S.~A.} \bibnamefont{Kivelson}},   
   \bibinfo{journal}{Phys. Rev. B}
 \textbf{\bibinfo{volume}{77}}, \bibinfo{pages}{224509} 
  (\bibinfo{year}{2008}).

\bibitem[{\citenamefont{Xu et~al.}(2008)\citenamefont{Xu, M\"uller, and Sachdev}}]{Sachdev}
    \bibinfo{author}{\bibfnamefont{C.} \bibnamefont{Xu}},
  \bibinfo{author}{\bibfnamefont{M.} \bibnamefont{M\"uller}},  \bibnamefont{and}
    \bibinfo{author}{\bibfnamefont{S.} \bibnamefont{Sachdev}},   
   \bibinfo{journal}{Phys. Rev. B}
 \textbf{\bibinfo{volume}{78}}, \bibinfo{pages}{020501(R)} 
  (\bibinfo{year}{2008}).

\bibitem[{\citenamefont{Chubukov et~al.}(2008)\citenamefont{Chubukov, Efremov, and Eremin}}]{Chubukov08}
    \bibinfo{author}{\bibfnamefont{A.~V.} \bibnamefont{Chubukov}},
  \bibinfo{author}{\bibfnamefont{D.~V.} \bibnamefont{Efremov}},  \bibnamefont{and}
    \bibinfo{author}{\bibfnamefont{I.} \bibnamefont{Eremin}},   
   \bibinfo{journal}{Phys. Rev. B}
 \textbf{\bibinfo{volume}{78}}, \bibinfo{pages}{134512} 
  (\bibinfo{year}{2008}).

\bibitem[{\citenamefont{Yamase et~al.}(2015)\citenamefont{Yamase, and Zeyher}}]{Yamase}
  \bibinfo{author}{\bibfnamefont{H.} \bibnamefont{Yamase}},  \bibnamefont{and}
    \bibinfo{author}{\bibfnamefont{R.} \bibnamefont{Zeyher}},  
   \bibinfo{journal}{arXiv:1503.07646}
  (\bibinfo{year}{2015}).

\bibitem[{\citenamefont{Devereaux et~al.}(2007)\citenamefont{Devereaux, and Hackl}}]{Rudireview}
    \bibinfo{author}{\bibfnamefont{T.~P.} \bibnamefont{Devereaux}},  \bibnamefont{and}
  \bibinfo{author}{\bibfnamefont{R.} \bibnamefont{Hackl}},
   \bibinfo{journal}{Rev. Mod. Phys.}
 \textbf{\bibinfo{volume}{79}}, \bibinfo{pages}{175} 
  (\bibinfo{year}{2007}).

\bibitem[{\citenamefont{Caprara et~al.}(2011)\citenamefont{Caprara, Di Castro, Muschler, Prestel, Hackl, Lambacher, Erb, Komiya, Ando, and Grilli}}]{Caprara11}
  \bibinfo{author}{\bibfnamefont{S.} \bibnamefont{Caprara}},
      \bibinfo{author}{\bibfnamefont{C.} \bibnamefont{Di Castro}},
    \bibinfo{author}{\bibfnamefont{B.} \bibnamefont{Muschler}},
    \bibinfo{author}{\bibfnamefont{W.} \bibnamefont{Prestel}},
    \bibinfo{author}{\bibfnamefont{R.} \bibnamefont{Hackl}},
    \bibinfo{author}{\bibfnamefont{M.} \bibnamefont{Lambacher}},
    \bibinfo{author}{\bibfnamefont{A.} \bibnamefont{Erb}},
    \bibinfo{author}{\bibfnamefont{S.} \bibnamefont{Komiya}},
    \bibinfo{author}{\bibfnamefont{Y.} \bibnamefont{Ando}},  \bibnamefont{and}
    \bibinfo{author}{\bibfnamefont{M.} \bibnamefont{Grilli}},
   \bibinfo{journal}{Phys. Rev. B}
 \textbf{\bibinfo{volume}{84}}, \bibinfo{pages}{054508} 
  (\bibinfo{year}{2011}).

\bibitem[{\citenamefont{Caprara et~al.}(2009)\citenamefont{Caprara, Di Castro, Enss, and Grilli}}]{Caprara09}
  \bibinfo{author}{\bibfnamefont{S.} \bibnamefont{Caprara}},
      \bibinfo{author}{\bibfnamefont{C.} \bibnamefont{Di Castro}},
    \bibinfo{author}{\bibfnamefont{T.} \bibnamefont{Enss}},  \bibnamefont{and}
    \bibinfo{author}{\bibfnamefont{M.} \bibnamefont{Grilli}},
   \bibinfo{journal}{Journal of Magnetism and Magnetic Materials}
 \textbf{\bibinfo{volume}{321}}, \bibinfo{pages}{686-689} 
  (\bibinfo{year}{2009}).

\bibitem[{\citenamefont{Tassini et~al.}(2005)\citenamefont{Tassini, Venturini, Zhang, Hackl, Kikugawa, and Fujita}}]{Tassini}
  \bibinfo{author}{\bibfnamefont{L.} \bibnamefont{Tassini}},
      \bibinfo{author}{\bibfnamefont{F.} \bibnamefont{Venturini}},
    \bibinfo{author}{\bibfnamefont{Q.~M.} \bibnamefont{Zhang}},
    \bibinfo{author}{\bibfnamefont{R.} \bibnamefont{Hackl}},
        \bibinfo{author}{\bibfnamefont{N.} \bibnamefont{Kikugawa}},  \bibnamefont{and}
        \bibinfo{author}{\bibfnamefont{T.} \bibnamefont{Fujita}},
   \bibinfo{journal}{Phys. Rev. Lett.}
 \textbf{\bibinfo{volume}{95}}, \bibinfo{pages}{117002} 
  (\bibinfo{year}{2005}).

\bibitem[{\citenamefont{Devereaux et~al.}(1999)\citenamefont{Devereaux, and Kampf}}]{Devereaux}
  \bibinfo{author}{\bibfnamefont{T.~P.} \bibnamefont{Devereaux}},  \bibnamefont{and}
      \bibinfo{author}{\bibfnamefont{A.~P.} \bibnamefont{Kampf}},
   \bibinfo{journal}{Phys. Rev. B}
 \textbf{\bibinfo{volume}{59}}, \bibinfo{pages}{6411} 
  (\bibinfo{year}{1999}).

\bibitem[{\citenamefont{Valenzuela et~al.}(2013)\citenamefont{Valenzuela, Calderon, Leon, and Bascones}}]{Valenzuela}
  \bibinfo{author}{\bibfnamefont{B.} \bibnamefont{Valenzuela}},
      \bibinfo{author}{\bibfnamefont{M.~J.} \bibnamefont{Calderon}},
    \bibinfo{author}{\bibfnamefont{G.} \bibnamefont{Leon}},  \bibnamefont{and}
    \bibinfo{author}{\bibfnamefont{E.} \bibnamefont{Bascones}},
   \bibinfo{journal}{Phys. Rev. B}
 \textbf{\bibinfo{volume}{87}}, \bibinfo{pages}{075136} 
  (\bibinfo{year}{2013}).

\bibitem[{\citenamefont{Paul}(2014)\citenamefont{Paul}}]{Indranil2}
  \bibinfo{author}{\bibfnamefont{I.} \bibnamefont{Paul}},
   \bibinfo{journal}{Phys. Rev. B}
 \textbf{\bibinfo{volume}{90}}, \bibinfo{pages}{115102} 
  (\bibinfo{year}{2014}).

\bibitem[{\citenamefont{Devereaux et~al.}(1994)\citenamefont{Devereaux, Einzel, Stadlober, Hackl, Leach, and Neumeier}}]{Devereauxdxy}
  \bibinfo{author}{\bibfnamefont{T.~P.} \bibnamefont{Devereaux}},
      \bibinfo{author}{\bibfnamefont{D.} \bibnamefont{Einzel}},
            \bibinfo{author}{\bibfnamefont{B.} \bibnamefont{Stadlober}},
 \bibinfo{author}{\bibfnamefont{R.} \bibnamefont{Hackl}},
   \bibinfo{author}{\bibfnamefont{D.~H.} \bibnamefont{Leach}},  \bibnamefont{and}
   \bibinfo{author}{\bibfnamefont{J.~J.} \bibnamefont{Neumeier}},
   \bibinfo{journal}{Phys. Rev. Lett.}
 \textbf{\bibinfo{volume}{72}}, \bibinfo{pages}{396} 
  (\bibinfo{year}{1994}).

\bibitem[{\citenamefont{Baek et~al.}(2015)\citenamefont{Baek, Efremov, Ok, Kim, van den Brink, and B\"uchner}}]{Baek}
  \bibinfo{author}{\bibfnamefont{S.~H.} \bibnamefont{Baek}},
    \bibinfo{author}{\bibfnamefont{D.~V.} \bibnamefont{Efremov}},
    \bibinfo{author}{\bibfnamefont{J.~M.} \bibnamefont{Ok}},
    \bibinfo{author}{\bibfnamefont{J.~S.} \bibnamefont{Kim}},
    \bibinfo{author}{\bibfnamefont{Jeroen} \bibnamefont{van den Brink}},  \bibnamefont{and}
        \bibinfo{author}{\bibfnamefont{B.~B.} \bibnamefont{B\"uchner}},
   \bibinfo{journal}{Nature Mat.}
 \textbf{\bibinfo{volume}{14}}, \bibinfo{pages}{210} 
  (\bibinfo{year}{2015}).

\bibitem[{\citenamefont{Azcarraga et~al.l}(1998)\citenamefont{de Azcarraga, Macfarlane, Mountain,  and Perez Bueno}}]{maths1}
  \bibinfo{author}{\bibfnamefont{J.~A.} \bibnamefont{de Azcarraga}},
    \bibinfo{author}{\bibfnamefont{A.~J.} \bibnamefont{Macfarlane}},
    \bibinfo{author}{\bibfnamefont{A.~J.} \bibnamefont{Mountain}},  \bibnamefont{and}
    \bibinfo{author}{\bibfnamefont{J.~C.} \bibnamefont{Perez Bueno}},
   \bibinfo{journal}{Nucl. Phys. B}
 \textbf{\bibinfo{volume}{510}}, \bibinfo{pages}{657-687} 
  (\bibinfo{year}{1998}).

\bibitem[{\citenamefont{Azcarraga et~al.l}(?)\citenamefont{de Azcarraga, and Macfarlane}}]{maths2}
  \bibinfo{author}{\bibfnamefont{J.~A.} \bibnamefont{de Azcarraga}},  \bibnamefont{and}
    \bibinfo{author}{\bibfnamefont{A.~J.} \bibnamefont{Macfarlane}},
   \bibinfo{journal}{IJMPA}
\textbf{\bibinfo{volume}{16}}, \bibinfo{pages}{1377-1405} 
  (\bibinfo{year}{2001}).

\end{thebibliography}
\end{document}